\documentclass{aa}
\usepackage{psfig}
\usepackage{epsfig}
\usepackage{graphics}
\usepackage{graphicx}
\usepackage[]{natbib}
\usepackage{txfonts}
\bibpunct[]{(}{)}{,}{a}{}{,}

\begin{document}

   \title{Flares from small to large: X-ray spectroscopy 
    of Proxima Centauri with XMM-Newton\thanks{Based 
          on observations obtained with XMM-Newton, an ESA science 
          mission with instruments and contributions directly funded by 
          ESA Member States and the USA (NASA)}  }


   \titlerunning{X-ray spectroscopy of Proxima Centauri}
   \authorrunning{M. G\"udel et al.}

   \author{Manuel G\"udel  
          \inst{1}
          \and
          Marc Audard
          \inst{2}
	  \and
          Fabio Reale          
	  \inst{3}
	  \and
          Stephen~L. Skinner 
	  \inst{4}
	  \and
          Jeffrey~L. Linsky          
	  \inst{5}
          }

   \offprints{M. G\"udel; \email{guedel@astro.phys.ethz.ch}}

   \institute{Paul Scherrer Institut, W\"urenlingen \& Villigen, CH-5232 Villigen PSI,
              Switzerland; \email{guedel@astro.phys.ethz.ch}
         \and
             Columbia Astrophysics Laboratory, Columbia University, 
             550 West 120th Street, New York,  NY 10027, USA; 
	     \email{audard@astro.columbia.edu}
         \and
             Dipartimento di Scienze Fisiche \& Astronomiche, Sezione di Astronomia,
	     Universit\`a di Palermo, Piazza del Parlamento 1, I-90134 Palermo, Italy; 
	     \email{reale@astropa.unipa.it} 
         \and
             Center for Astrophysics and Space Astronomy, University of Colorado, 
	     Boulder, CO 80309-0389, USA; 
	     \email{skinners@casa.colorado.edu}
         \and 
	     JILA, University of Colorado and NIST, Boulder, CO 80309-0440, USA; 
             \email{jlinsky@jila.colorado.edu}    }
	     
   \date{Received $<$date$>$; accepted $<$date$>$}

\abstract{We report results from a comprehensive study of the nearby M dwarf Proxima
Centauri with the XMM-Newton satellite, using simultaneously its X-ray detectors and
the Optical Monitor with its U band filter.
We find strongly variable coronal X-ray emission, with  flares ranging over a factor of 100
in peak flux.  The low-level emission is found to be continuously variable on at least three time scales
(a slow decay of several hours, modulation on a time scale of 1 hr, and weak flares with
time scales of a few minutes). Several weak flares are characteristically preceded by an 
optical burst, compatible with predictions from standard solar flare models.
We propose that the U band bursts are proxies 
for the elusive stellar non-thermal hard X-ray bursts  suggested from solar observations.  
In the course of the observation, a very large X-ray flare started and was observed essentially 
in its entirety. Its peak luminosity reached $3.9\times 10^{28}$~erg~s$^{-1}$
[0.15--10~keV], and the total X-ray energy released in the same band is derived to be $1.5\times
10^{32}$~ergs. This flare has for the first time allowed to measure significant density variations
across several phases of the flare from X-ray spectroscopy of the O\,{\sc vii} He-like triplet; we find peak 
densities reaching up to $4\times 10^{11}$~cm$^{-3}$ for plasma of about $1-5$~MK. Abundance 
ratios show little variability in time, with a tendency of
elements with a high first ionization potential to be overabundant relative to solar photospheric
values. Using Fe\,{\sc xvii} lines with different oscillator strengths, we do not find significant
effects due to opacity during the flare, indicating that large opacity increases are not the 
rule even in extreme flares. 
We model the large flare in terms of an analytic 2-Ribbon flare model and find that the 
flaring loop system should have large characteristic sizes ($\approx 1R_*$) within the framework of 
this simplistic model. These results are supported by full hydrodynamic simulations. 
Comparing the large flare to flares of similar
size occurring much more frequently on more active stars, we propose that the X-ray properties
of active stars are a consequence of superimposed flares such as the example 
analyzed in this paper. Since 
larger flares produce hotter plasma, such a model also explains why, during episodes of
low-level emission, more active stars show hotter plasma than less active stars.

}
\maketitle

%

\section{Introduction}

Coronal energy release on magnetically active cool stars is, according to current
understanding, largely driven by the interplay of surface convective motion
and magnetic fields anchored in the star's outer convection zone. Strong  
variability in coronal emissions due to flares 
(in X-rays, the extreme ultraviolet, or the radio range) support theories
proposing magnetic instabilities as the main drivers of energy release and
transport (see \citealt{haisch91} for a general multi-wavelength review of solar and
stellar flares). Although stellar coronal radiation has traditionally been
classified as ``quiescent'' or ``flare'' depending on the recorded
time scale of variability, recent studies have cast some doubt on
the overall meaning of this distinction for magnetically very
active main-sequence stars. First, sensitivity limitations inevitably smooth out low-level 
variability. Second, a superposition
of many strongly overlapping flares tends to produce smooth light curves
in particular for statistical distributions of flare energies for which
the low-energy population dominates the total energy budget \citep{kopp93, guedel00}.
And third, long time series monitoring stellar X-ray or EUV
activity have provided evidence for continuous variability ascribed to flares,
to an extent that possibly all of the observed coronal emission can be attributed
to explosive energy release \citep{audard00, kashyap02, guedel03a}. This latter
hypothesis is presently also in the center of solar coronal research (e.g.,
\citealt{porter95, krucker98, parnell00, aschwanden00}) in the context of
``micro-'' or ``nano-''\-flares. 

Large, outstanding coronal flares may thus be the prototypes of the coronal heating
agents accessible relatively easily by monitoring programs available 
on X-ray satellites. Their investigation may provide the fundamental parameters
required to understand explosive energy release by magnetic reconnection,
shed light on the magnetic geometry prevailing in stellar flares, and
help understand processes transporting chromospheric and photospheric
material into the corona. Comparing such studies on magnetically active
stars with solar findings is particularly important as magnetic processes on
the former may fundamentally differ from those on the Sun, given the
much larger filling factor of surface magnetic fields on active stars.
{\it XMM-Newton} and
{\it Chandra} offer a new dimension to stellar flare research by providing
access to highly sensitive high-resolution X-ray spectroscopy, thus
potentially revealing information not only on the temperature structure but
also on electron  densities, turbulence, or bulk mass motions.

During a recent {\it XMM-Newton} campaign on the nearest star to the Sun, 
Proxima Centauri, we were fortunate in recording a major long-duration X-ray 
and optical flare that outshone the average pre-flare emission by almost two orders 
of magnitude in X-rays, and several magnitudes in the optical. Early findings
from this campaign are described in G\"udel et al. (2002a = Paper I). The observing program
was initially devised to investigate the statistics of low-level activity,
providing access to the weakest flares yet on any star other than the Sun.
The campaign indeed recorded a sequence of low-level flares to an extent
that no part of the light curve can be described as steady, for a duration of
about 35~ks (Paper I).  Many X-ray events were characteristically 
preceded by optical bursts in a manner predicted by the chromospheric evaporation 
scenario \citep{antonucci84, dennis88, hudson92}. The large flare for the first time 
provided a sufficient statistical signal-to-noise
ratio to explicitly and significantly measure the run of electron densities
in the flare plasma from spectroscopic He-like line triplets. Electron
densities of several times $10^{11}$~cm$^{-3}$ were found during peak
time, while they rapidly decreased during the flare decay phase.
These observations compare favorably with previous solar spectroscopic measurements
of flare densities \citep{mckenzie80, doschek81}.

The present paper presents the complete {\it XMM-Newton} data set on Proxima 
Centauri and discusses some models for the large flare. In a companion paper \citep{reale04},
we will present a series of hydrodynamic simulations for the same flare.
The paper is structured as follows:  Sections 2 and 3 describe the target and the 
observations, Sect. 4 presents the analysis and our results, while Sect. 5 discusses
the findings in the context of flare physics and coronal heating. Sect. 6 contains
our conclusions.
  
\section{The Target}

Proxima Centauri is a magnetically active dM5.5e dwarf ($V = 11.05$, see SIMBAD database;
$L_{\rm bol} = 6.7\times 10^{30}$~ergs$^{-1}$, see \citealt{frogel72})
revealing strong coronal activity. At a distance of $1.295\pm 0.005$~pc
\citep{perryman97}, it is the closest star to the Sun.
Its ``quiescent'' X-ray luminosity $L_{\rm X} \approx (4-16)\times 10^{26}$~erg~s$^{-1}$ 
\citep{haisch90} is
similar to the Sun's  despite its $\approx$50 times smaller surface area. It 
has attracted the attention of most previous X-ray observatories. \citet{haisch80}
and \citet{haisch81} observed quiescent and flaring X-ray emission from 
this star with the {\it Einstein} satellite, with the largest flare reaching an 
X-ray luminosity of $L_{\rm X} = 7.4\times 10^{27}$~erg~s$^{-1}$ and peak temperatures 
of 17~MK during the rise to the luminosity peak. The surface filling factor of X-ray 
emitting magnetically confined plasma was estimated to be about 20\% based on solar analogy.
Interestingly, coordinated radio, optical, and ultraviolet observations recorded
no flare emissions. The authors interpreted this long-decay flare (decay time
scale $\tau \approx 20$~min) by a series of arches cooling predominantly
by radiation, with dimensions comparable to the stellar radius. In a follow-up observation,
\citet{haisch83} report another very large flare, with a peak X-ray 
luminosity of $L_{\rm X} = 1.4\times 10^{28}$~erg~s$^{-1}$ in the 0.2--4~keV 
band, and a maximum temperature of 27~MK. A two-ribbon flare model in
analogy to large solar flares was proposed for this long-duration event 
($\tau \approx 20$~min), refined later by \citet{poletto88}. Further modeling of this 
flare, involving hydrodynamic simulations of flaring loops, were presented by 
\citet{reale88}. Both latter references found flaring loop structures with sizes up to
about $7\times 10^9$~cm or 0.7$R_*$.

\citet{haisch90} present a coordinated {\it EXOSAT}-{\it IUE} observation
of Proxima Centauri, again comprising one larger flare (peak $L_{\rm X} = 3\times 10^{27}$~erg~s$^{-1}$). 
\citet{haisch95} obtained rather sensitive observations
of Proxima Centauri in the harder bands provided by the {\it ASCA} detectors
where the contrast between (hotter) flares and the (cooler) quasi-steady
emission is more pronounced. The authors point out that a large number
of intermediate flares seem to be present in the light curve, corresponding
to solar M-class flares. Even harder emission was detected by the {\it XTE}
satellite, but unfortunately no large flares occurred during the time
of the observations \citep{haisch98}.

Proxima Centauri is an obvious target also for the newest generation of 
satellites, {\it XMM-Newton} and {\it Chandra}. First results of the
observation presented here have been reported in Paper I 
and \citet{guedel03b} pertaining to flare density increases, 
the Neupert effect \citep{neupert68} for X-ray and optical flares, and continuous low-level
flaring. \citet{wargelin02} discuss a CCD observation obtained by
{\it Chandra} concentrating on  the possibility to detect
charge exchange from the stellar wind.

\section{Observations}

{\it XMM-Newton} \citep{jansen01} observed Proxima Centauri on 2001 August 12 during 65~ks of 
exposure time.  All detectors were operating nominally. The observing log is
given in Table~\ref{obslog}. The data were reduced using
the Standard Analysis System (SAS), version 5.3 for the EPIC data \citep{strueder01, turner01} and
version 5.4.1 for the RGS spectra \citep{herder01}. 

In anticipation of possible large
flares occurring on this star, we used the small window modes for all EPIC detectors, 
avoiding X-ray pile-up effects up to count rates of about 5~cts~s$^{-1}$  for the two MOS detectors 
and up to 130~cts~s$^{-1}$  for the PN camera. The large flare reported here exceeds these limits
for the MOS but not for the PN. We will thus use only the PN for CCD spectral
analysis but combine data from all three detectors to produce sensitive light curves
during the low-level emission.

We extracted the EPIC spectral 
data from circles with a radius of 50$^{\prime\prime}$ (encircling 90\% of the source photons). 
For the  PN, the corresponding response matrix and the effective area were derived using the 
standard SAS tasks rmfgen and arfgen. The energy range of sensitivity of the EPIC cameras 
is 0.1--12~keV.  Their energy resolving power is best at highest 
energies;  it reaches $E/\Delta E \approx 50$ at 7~keV for the PN and scales as $E^{1/2}$. 

Background light curves were extracted from  CCD fields outside the source region (for the 
MOS, we used four background areas in the outer chips for improved statistics; background 
vignetting is negligible since the background ist dominated by intrinsic detector noise
and by soft proton irradiation; both components are only weakly affected by
vignetting; we refer to the XMM-Newton Users Handbook, \S~3.3.7 for further details). 
A few small background flares induced by proton events in the CCD were seen, with
amplitudes in the combined EPIC PN+MOS1+MOS2 data (see below) of no more than 0.65~ct~s$^{-1}$ during the first 1.7~hrs, 
and of no  more than 0.33~ct~s$^{-1}$ thereafter. These events were reliably eliminated
by subtracting the background light curve. The median background rate was relatively high,
with 0.08~cts~s$^{-1}$ in the combined light curve.


   \begin{table}[b!]
      \caption[]{Observing log\label{obslog}}
      \[
         \begin{array}{lllll}
            \hline
            \hline
           \noalign{\smallskip}
            &  & \multicolumn{3}{l}{\rm 2001\ August\ 12}\\
            \noalign{\smallskip}
            \hline
            \noalign{\smallskip}
            {\rm Instrument}  &  & {\rm UT\ range}   &  & {\rm JD\  2452133.5 +}  \\
            \noalign{\smallskip}
            \hline
            \noalign{\smallskip}
            {\rm PN}     &\quad   &  04:21:29 - 22:37:29 &\quad & 0.18147 - 0.94270\\
            {\rm MOS1}   &\quad   &  04:01:29 - 22:35:08 &\quad & 0.16770 - 0.94106\\
            {\rm MOS2}   &\quad   &  04:01:29 - 22:35:08 &\quad & 0.16770 - 0.94106\\
            {\rm RGS1}   &\quad   &  03:53:59 - 22:37:34 &\quad & 0.16249 - 0.94275\\
            {\rm RGS2}   &\quad   &  03:53:59 - 22:37:32 &\quad & 0.16249 - 0.94273\\
            {\rm OM}     &\quad   &  03:58:58 - 17:43:46 &\quad & 0.16595 - 0.73873\\
           \noalign{\smallskip}
            \hline
         \end{array}
      \]\normalsize
      \vskip -0.6truecm
   \end{table}

The two Reflection Grating Spectrometers (RGS) provide 
high-resolution spectroscopy ($E/\Delta E \approx 300$ FWHM around the O\,{\sc vii} 
lines at $\approx  22$~\AA) over the wavelength range of 5-35~\AA, with a maximum first order
effective area of about 120~cm$^2$ at 15~\AA. To include recent updates of
the effective area function, we reanalyzed the RGS data using the SAS 5.4.1 release but
note that there is no significant change with respect to the results reported previously
(Paper I, using SAS vers. 5.3.3 for the data reduction). The background was determined 
from adjacent areas on the RGS CCD detectors as prescribed in the SAS task rgsproc. 
The response matrix (including the effective  area function) was produced with the SAS 
task rgsrmfgen.

The Optical Monitor (OM;  \citealt{mason01})  observed the star through its U band 
filter in high-time resolution mode. The data reduction used the updates provided by
SAS vers. 5.3.3. The count rates observed during the 
flares approach saturation of the detector. As the true photon flux must
be estimated based on detector dead-time corrections, the photon count rate
numbers given in this paper are somewhat uncertain and are based on the
current understanding of the calibration. Given the uncertain physical nature
of the optical emission, the level of accuracy is entirely adequate for
our purposes.

\section{Data analysis and results}

\subsection{Light curves}\label{lightc}

The light curves extracted from the EPIC cameras are shown in Fig. 1 and 2.
The first $\approx$12 hours of the observing time recorded low level radiation.
This portion of the observation is shown in Fig. 1.
At $t \approx 0.715$~d, an extremely large flare sets in, rising
to its luminosity peak in about 12 minutes. A very long decay follows, with several 
superimposed secondary flares visible. This episode is illustrated in Fig. 2. 

\begin{figure*} 
\includegraphics[width=16cm]{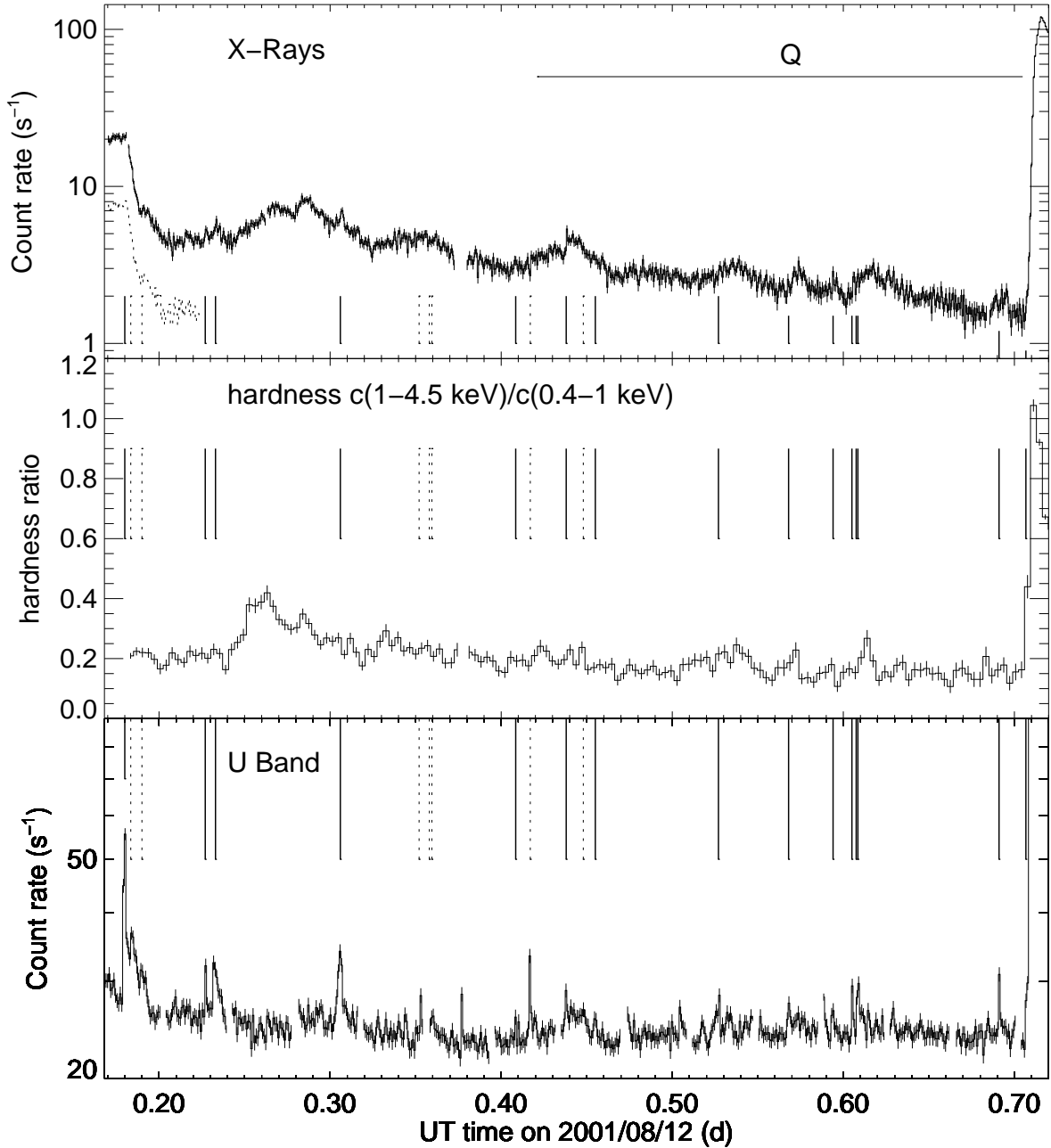}
\caption{Low level emission of Proxima Centauri. 
{\it Top panel:} X-ray light curve. Counts from all three EPIC detectors have been
co-added (using counts between 0.18--4.5~keV from the MOS detectors, and
0.15--4.5~keV from the PN). Time resolution is 60~s. The horizontal bar marks the interval
that was used for spectral analysis of low-level emission. The vertical lines (also in
the other panels) indicate flare-like features. Solid lines refer to clearly detected
features (in the X-ray or the optical band), while dotted lines mark borderline cases. 
The dashed curve until 0.22~d is the MOS-only light curve
shown here because the PN was not in operation before $\approx 0.18$~d (the total light
curve has been approximately scaled). 
{\it Middle panel:} Hardness, defined as the ratio of counts between 1--4.5~keV to counts 
between 0.4--1~keV, at a time resolution of 300~s. Again, counts from all EPIC detectors were
co-added.
{\it Bottom panel:} Optical U-band light curve from OM (on a logarithmic count rate scale). The 
time resolution is 80~sec per bin. Time gaps are of instrumental origin, specific for the selected 
observing mode.
\label{light1}}
\end{figure*}

\begin{figure*} 
\includegraphics[width=16cm]{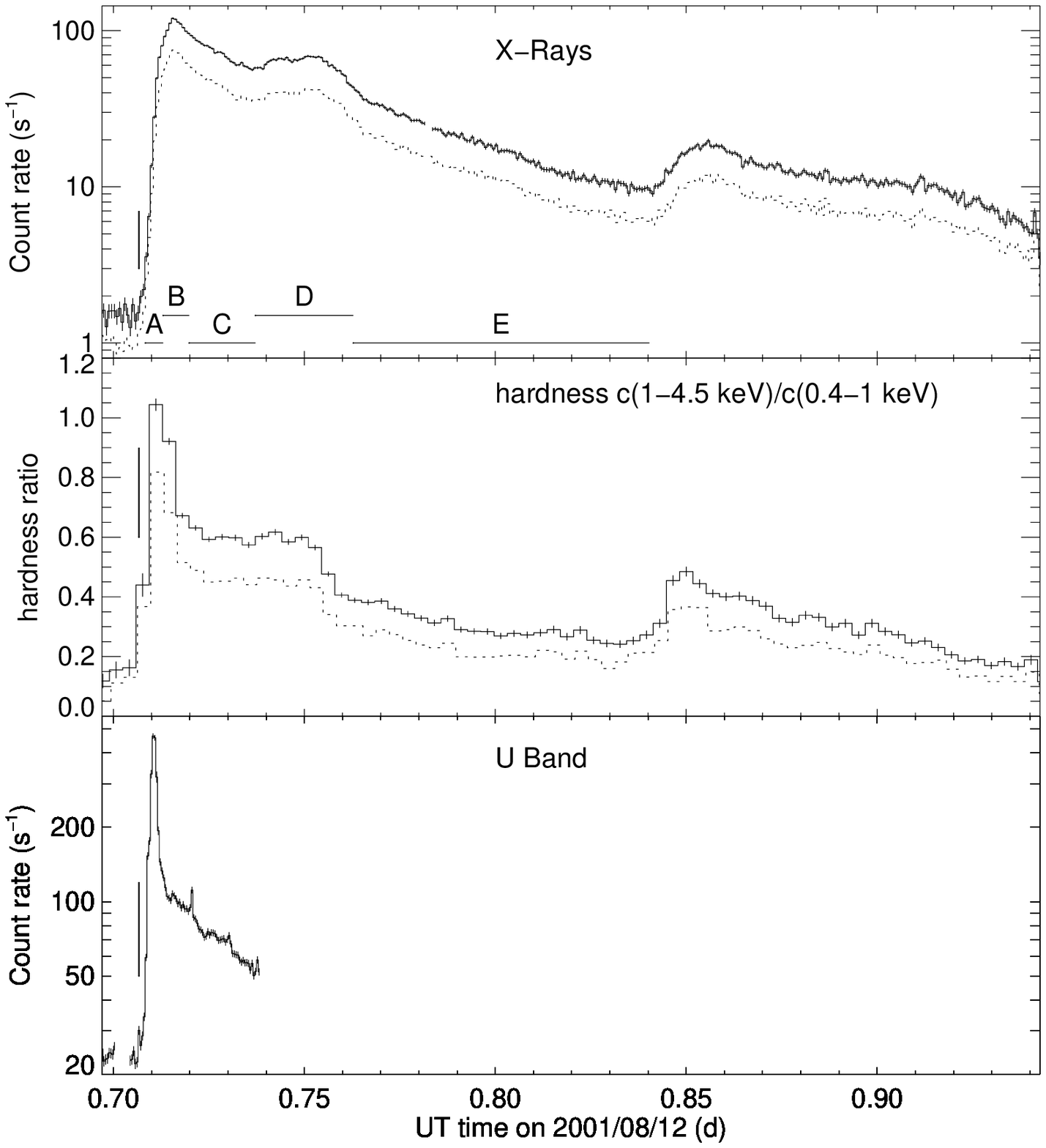}
\caption{Similar to Fig.~\ref{light1}, with the following exceptions:
The top and the middle panels additionally show the PN-only curves (dashed), since only
those data were not piled up during the flare peak. And the optical light curve in the bottom panel
is shown at a time resolution of 40~s, to illustrate some finer details during high flux. 
The OM observation was terminated early, but covers the main flare peak. The intervals
A--E used for spectroscopy in this paper are marked in the top panel.
\label{light2}}
\end{figure*}

For our later analysis, we define time sections of
i) pre-flare low level emission (Q) ii) flare rise (A), iii) first flare peak (B), 
iii) initial flare decay (C), 
iv) secondary flare peak (D), and iv) final flare decay (E), as shown in Figs.~\ref{light1} and
\ref{light2}.

Fig. 1 shows, from top to bottom, the summed EPIC PN+MOS1+MOS2 count rate at a resolution 
of 60~s. We have confined our photon extraction to the energy intervals 0.15--4.5~keV for
PN and 0.18-4.5~keV for MOS in order to suppress contaminating warm pixel contributions and
excessive background, while optimizing the signal-to-noise ratio. Tests with lower energy limits 
up to 0.4~keV did not show significantly different results. The initial $\approx$20
min (until 0.1815~d, see Table~\ref{obslog}) were not observed  by the PN camera; we multiplied 
the MOS count rates by 2.7 to obtain an estimate for the total PN+MOS light curve for that interval but 
note that, due to the spectral dependence of the PN/MOS count rate ratio, this 
estimate is only approximative and the absolute count rate values before 0.1815~d
should be accepted with some reservations. We therefore also plot the MOS-only light curve
for the early time interval (dashed in Fig.~\ref{light1}). The middle panel in Fig. 1 
presents the hardness ratio at a resolution of 300~s, here defined as the ratio between 
the count rates in the  1--4.5~keV and  the 0.4--1~keV bands. 
Hardness is thus a sensitive indicator of heating events. The lowest panel shows the OM 
light curve at a resolution of  80~s per bin, chosen for optimum illustration of 
rapid flares.

The most surprising aspect of this part of the observation is the level of continuous
variability. At no time is the light curve constant; even if the long-term trend is
removed, there are hardly any intervals of more than a few tens of minutes that are free
of modulations at the given sensitivity.  This is providing evidence of continuous 
{\it microvariability.} There are three obvious time scales of variability:

i) The X-ray light curve slowly decays by an order of magnitude from an initial 
$\approx 10$~cts~s$^{-1}$ to $\approx 1.5$~cts~s$^{-1}$ at $t = 0.7$~d.  
The count rate-to-luminosity conversion factor applicable to this 
part of the light curve has been derived from spectral modeling and amounts to $\approx 
2.8\times 10^{26}$~erg~s$^{-1}$ per ct~s$^{-1}$. From this, the low-level luminosity  
varies between
$\approx (4-28) \times 10^{26}$~erg~s$^{-1}$, compatible with previous literature values.
The cause of the long-term decay is unclear. It could be due to rotational modulation of
a stellar surface that is inhomogeneously covered  by X-ray emitting active regions, most of 
which rotate out of sight during the observation. The rotation period of Proxima Centauri
is, however,  suspected to be as long as 83.5~d, see \citet{benedict98}. Alternatively, 
the slowly decaying emission may be part of the late light curve of a giant flare occurring 
mostly before our observations, similar to the flare occurring during the second part of our 
observations.

ii) A shorter time scale of about 1~hr reveals itself as a slow modulation in the X-ray 
light curve. At closer inspection, some of gradual peaks are in fact composed of
superimposed weak flares, as was shown in Paper I. They often correspond to 
groups of simultaneously occurring optical flares. A particularly clear case is seen at
$t = 0.43-0.46$~d in Fig. 1. The hardness light curve mostly shows enhancements
during the increasing portions of the X-ray light curve on the same time scale.
This indicates increased flare activity during the rise phase and decreased activity during
the decaying phase when plasma cools and the emission becomes softer.

iii) We  eliminated the above modulations from the light curve by subtracting
a smoothed version of the curve, leaving only short-term modulations in the
residuals. The autocorrelation function of this ``high-pass'' filtered curve
shows a decay from unity at zero lag down to zero correlation at about 4 minutes time lag. 
This is not compatible with residual noise but in fact corresponds to the typical time 
scales of the weakest flares, some of which were illustrated in 
Paper I. These flares  have peak luminosities of  $(2-5)\times 10^{26}$~erg~s$^{-1}$,
corresponding to total released soft X-ray energies of  $\approx 10^{28}-10^{29}$~erg 
(in the 0.1-10~keV range). Such flares would be classified as ``M-class flares'' on the Sun
(see \citealt{haisch95} for a discussion of M-class flares in the context of stellar
observations). Many of the detected X-ray flares are characteristically accompanied by
optical flares which typically occur during the rise time of the X-ray flare and
provide support for a flare model in which accelerated electrons (signified by
the optical burst as they collide in the chromosphere; \citealt{hudson92}) deposit their energy in the
chromosphere, heating and evaporating the plasma that subsequently fills the magnetic loops
(see, e.g., \citealt{antonucci84, dennis88}).
Such ``chromospheric evaporation'' appears to occur at a high cadence in Proxima
Centauri; clear examples are marked by solid vertical bars in all three panels in Fig. 1,
somewhat less obvious cases by dotted bars. Most of these events have been described
in Paper I; we refer to that paper for a more detailed discussion on low-level
flaring. Some of the outstanding 
examples can also be traced in the hardness curve although the dominating continuous or 
slowly varying  level of radiation overwhelms short-term hardness changes in the 
superimposed flares.

Fig. 2 shows the large flare, in three panels equivalent to those described in Fig. 1,
except that we plot the optical light curve with a time resolution of 40~s to bring out
some finer details during the high-count rate episode.  
The Proxima Centauri source was partly piled up in the MOS detectors  
during the flare, reaching $\approx$23~cts~s$^{-1}$ each at flare peak, while
the upper limit to avoid CCD pileup is $\approx 5$~cts~s$^{-1}$ for the MOS small window mode
(see XMM-Newton Users' Handbook, Sect. 3.3). The pile-up limit for the PN small window
is 130~ct~s$^{-1}$, while our flare peaks at $\approx 78$~ct~s$^{-1}$. 
We therefore also plot the count rate and hardness curves for data extracted from
the PN detector only. 
The total count rate and the hardness curves appear to be  little affected by
MOS pileup, however, as the comparison between the light curve pairs shows. The appropriate 
count rate-to-luminosity conversion factor for the PN-only data is 
$4.95\times 10^{26}$~erg~s$^{-1}$ per ct~s$^{-1}$ at flare peak. The ratio changes only
within a few percent from the flare peak to the decay phase. The peak luminosity of this
flare is thus $\approx 3.9\times 10^{28}$~erg~s$^{-1}$ [0.15-10~keV], and the total energy emitted
in X-rays is of order $1.5\times 10^{32}$~erg.
As predicted by the chromospheric evaporation scenario, hardness peaks during the rise of
the X-ray light curve, i.e., the highest temperatures are attained before the emission
measure has built up to its maximum. Similarly, the optical light curve peaks during
the most rapid increase of the X-ray curve. 
We also note that both the X-ray and the optical light curves
show a very faint precursor flare, marked by a vertical line, and again it is the optical 
signal that precedes the X-ray flare. The slowly decaying light curve shows two large,
superimposed events, one at 0.75~d and one at 0.85~d. Unfortunately, the OM instrument
did not continue to observe sufficiently long to cover these two events as its monitoring
program was terminated prematurely for operational reasons. They differ
in that the earlier one does not result in a significant increase of the hardness
(but rather in a plateau of constant hardness), while the later flare does. This may
be the result of ``hardness contrast'': When the later secondary flare occurs, the
hardness of the decaying initial flare plasma has already decayed nearly to levels
seen in Fig. 1. Further faint flares can be seen during the 
decay phase, such as the one at 0.91~d. They seem to correspond to ongoing low-level
flaring as detected in Fig. 1.


\subsection{Spectroscopy with EPIC and RGS}

\begin{figure} 
\includegraphics[width=8.8cm]{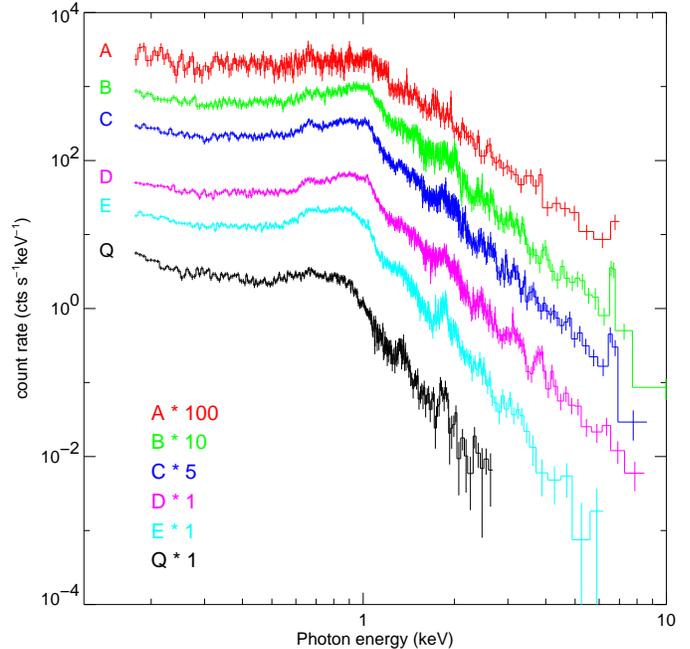}
\caption{EPIC PN spectra during various stages of the flare (A--E) and during the pre-flare
episode Q. The count rates have been multiplied by various factors to prevent
mutual overlap (see factors indicated at lower left). Note the variable high-energy slope,
indicative of variable flare temperatures.\label{epic}}
\end{figure}

The EPIC PN spectra shown in Fig.~\ref{epic} characterize the spectral evolution of the 
large flare across episodes A -- E (together with the low-level spectrum 'Q'). 
The individual spectra have been shifted along 
the y axis by factors indicated in the figure to avoid mutual overlap. The high-energy 
slope (between 1.5--7~keV) is largely determined by the bremsstrahlung continuum of
the hottest dominant plasma component. As the flare progresses, the slope becomes
steeper, indicative of overall cooling of the flare plasma.  Also, a prominent Fe K 
complex including He-like Fe\,{\sc xxv} lines is seen around 6.7~keV predominantly in the earlier 
flare phases and at flare
peak. Again, these lines are indicative of the presence of very hot ($> 10$~MK) plasma.
And third, the Fe L-shell region around 0.6--1.0~keV is another characteristic
indicator of temperature since it contains Fe lines with ionization stages between
Fe\,{\sc xvii}--Fe\,{\sc xxiv} that shift the peak emission of the unresolved spectral maximum from
$\approx 0.75$~keV in the pre-flare spectrum to $\approx$1~keV at flare peak, decreasing again
as the flare progresses.
 Various line features above 1.4~keV can be used to constrain abundances of Mg, Si, S, Ar and
Ca. We will present an abundance analysis below, combining EPIC information with RGS spectroscopy.

\begin{figure*} 
\includegraphics[width=16.cm]{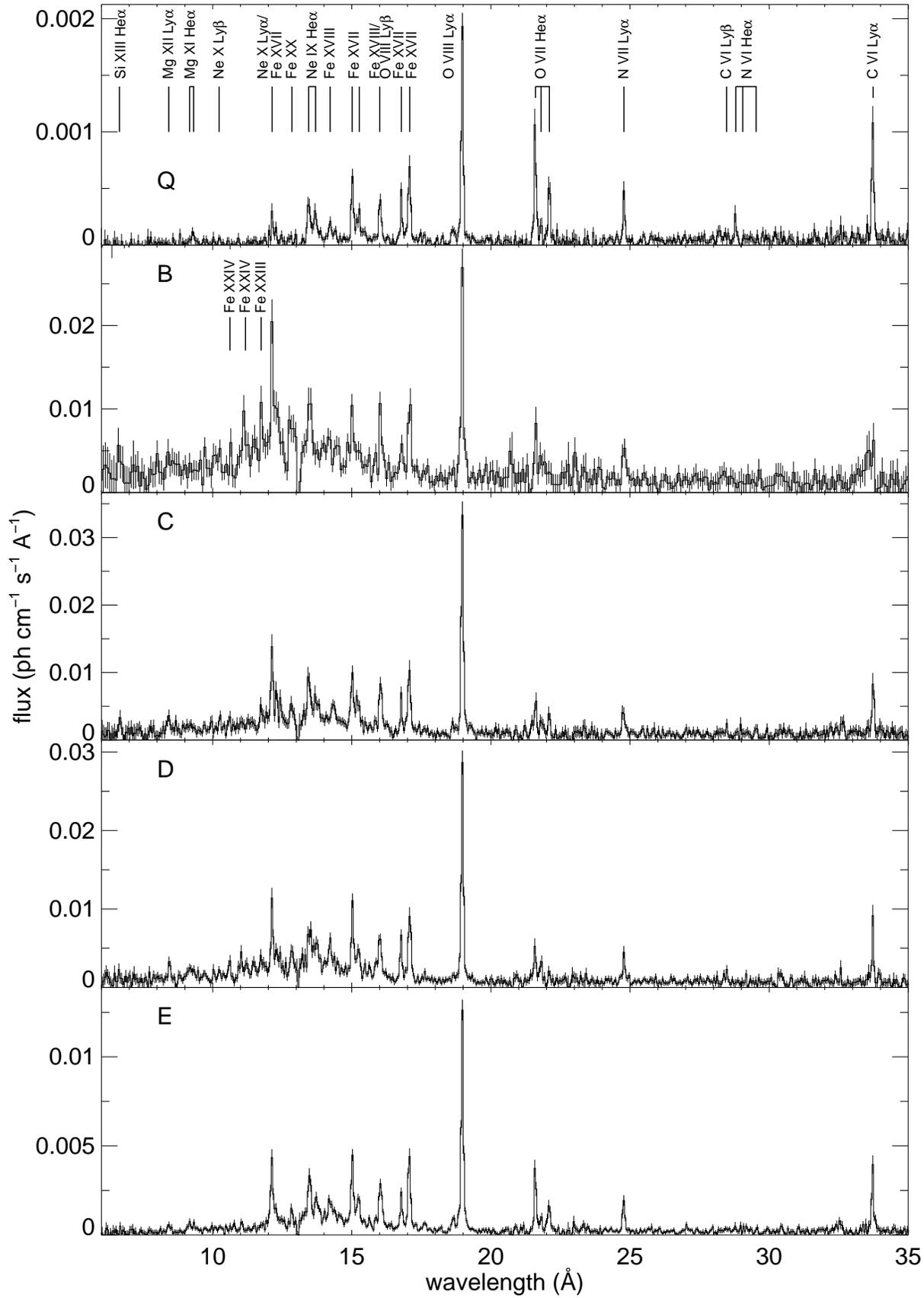}
\caption{Fluxed RGS1+2 spectra referring to the intervals Q (pre-flare low level),
and B--E (see Fig.~\ref{light2} for reference). The bin resolution is 50~m\AA\ except
for panel B where a resolution of 78~m\AA\ has been used to improve the signal-to-noise
ratio. 
\label{spectra}}
\end{figure*}

Reflection grating spectra obtained during the low-level episode before the flare and during four 
intervals from flare peak to late decay are shown in Fig.~\ref{spectra}. (The spectrum
obtained from interval A shows a very low signal-to-noise ratio; the analysis pertaining
to this interval strongly relies on the EPIC PN spectrum.) A few important
lines are marked. The spectra obtained during low-level emission (interval Q) and at  
flare peak (interval B)
are strikingly different, indicating heating to high temperatures during the flare.
For example, a marked  enhancement of the continuum flux level is visible at flare
peak, and the appearance of lines of Fe\,{\sc xx}--Fe\,{\sc xxiv} around 10--12~\AA\ with
peak formation temperatures between 10--20~MK
indicate the presence of large amounts of hot plasma.  

The spectra were used to study emission measure (EM) distributions, abundances, densities,
and optical depths at various stages of the flare development. We discuss
results in the following subsections.
 
\subsection{Spectroscopic density measurements}\label{densovii}
  
The He-like line triplets of O\,{\sc vii} (at 21.6--22.1~\AA) and of Ne\,{\sc ix} (at 13.4--13.7~\AA)
can be used to obtain characteristic electron densities in the source
region   of the respective lines \citep{gabriel69}. The ratio of the fluxes
of the forbidden line ($f$, $1s^2~^1S_0 - 1s2s~^3S_1$, at 22.1~\AA\ for O\,{\sc vii}
and at 13.7~\AA\ for Ne\,{\sc ix}) and the intercombination line ($i$, $1s^2~^1S_0 - 1s2p~^3P_{1,2}$,
at 21.8~\AA\ for O\,{\sc vii} and at 13.55~\AA\ for Ne\,{\sc ix}) are sensitive to the electron 
density $n_e$ in the range of $10^9-10^{12}$~cm$^{-3}$ for O\,{\sc vii} and  $10^{10}-10^{13}$~cm$^{-3}$  for
Ne\,{\sc ix} \citep{porquet01}. Further He-like triplets, in particular of Mg\,{\sc xi} and Si\,{\sc xiii}, are accessible within
the RGS range, but the spectral resolution is insufficient to cleanly separate 
the the lines of the triplet, and the density-sensitive ranges ($>5\times 10^{11}$~cm$^{-3}$)
are probably only marginally useful for typical coronal flare plasmas on dwarf stars.

\begin{figure*} 
\hbox{\hskip -1truecm\includegraphics[width=18.8cm]{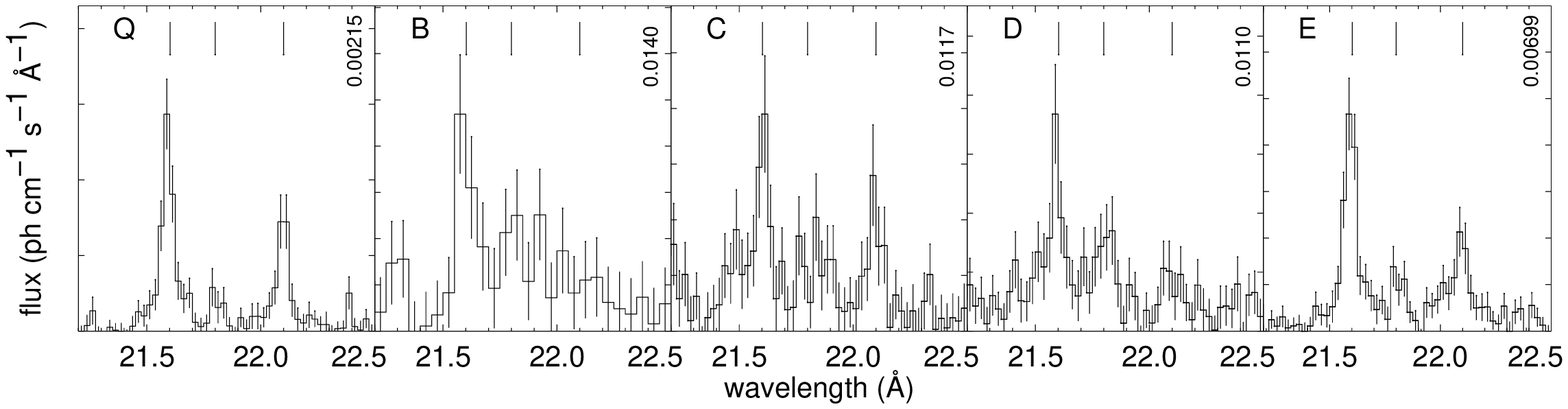}}
\caption{O\,{\sc vii} triplets during the low-level episode Q and during various flare intervals.
The small number at upper right indicates the flux at the upper border of the plot.
The three vertical lines mark the positions of the resonance, the intercombination, and the
forbidden line (for increasing wavelength). 
\label{ovii}}
\end{figure*}

In Paper I, we discussed the evolution of the $f/i$ ratios during the large flare episode;
we will not repeat the analysis but briefly summarize the results. The densities
derived from O\,{\sc vii} flux ratios were found to peak around $4\times 10^{11}$~cm$^{-3}$
during the primary and secondary flare peaks (intervals B and D, respectively), while
$n_e$ was lower by about one order of magnitude during the decay episodes (intervals
C and E). These densities combined with the spectral measurement of the EMs
of the O\,{\sc vii} line forming plasma were used to estimate total masses and volumes 
during the flare. It was found that the mass of the relatively cool O\,{\sc vii} forming
plasma increased appreciably toward the decay episodes, possibly due to continuous cooling
of plasma that was initially heated to much higher temperatures. The O\,{\sc vii} line triplets
are represented in flux units in Fig.~\ref{ovii} (B--E). Suppression of the $f$ line in panels
B and D is clearly visible. The spectrum obtained during the low-level episode (Q)
shows a line ratio compatible with fluxes around the lower limit of the density-sensitive 
range of the O\,{\sc vii} triplet, corresponding to densities in the range $10^9-10^{10}$~cm$^{-3}$
(see Paper I). 

\begin{figure*} 
\hbox{\hskip -1truecm\includegraphics[width=18.8cm]{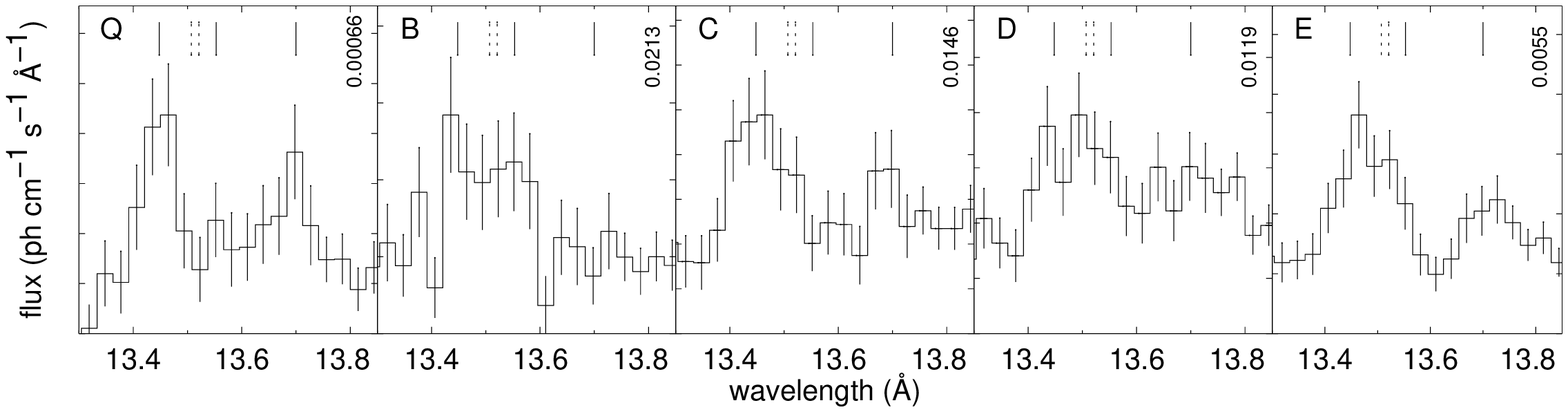}}
\caption{Same as Fig.~\ref{ovii}, but for the Ne\,{\sc ix} triplet. The dashed lines
indicate two potentially strong, blending Fe\,{\sc xix} lines.\label{neix}}
\end{figure*}

The analysis of the Ne\,{\sc ix} triplet (see Fig.~\ref{neix} for an illustration of fluxed spectral extractions)
is much more complicated due to strong blending with several lines in particular of
Fe\,{\sc xix} (\citealt{ness03b}, Fe\,{\sc xix} has a peak formation temperature of about 8~MK, 
compared to 4~MK for Ne\,{\sc ix}). While Fig.~\ref{neix}C and E do show a flux peak at the position
of the $f$ line at 13.7~\AA, its possible suppression in panels B and D cannot be
established given the dominance of contributions from Fe\,{\sc xix} at flare peak.

In principle, there may be some contributions to the $i$ and $f$ lines from
the steady component; the $f/i$ ratios may thus be slightly biased by non-flare
contributions. Since the $f$ line is much stronger than the $i$ line during the low-level 
emission (Fig.~\ref{ovii}Q), the correction is essentially for the former flux. 
If the decaying trend of the overall light curve (Fig.~\ref{light1}) continues after the start
of the large flare, then, for the intervals D--E, we expect an O\,{\sc vii} flux about 2.5 
times smaller than on average during the low-level interval Q, resulting in a
non-flare $f$ flux about 13\% of
that observed during  interval D. Correction of this effect would increase the density
by about 15\%, which is not significant given the considerable errors in the flux ratios.
In any event, we note that the densities reported in Paper I are in this sense lower limits, but 
specific corrections are meaningless since we cannot accurately estimate the (small) contribution 
of the non-flare emission at any time during our strong flare. The effect of low-level contributions
of course becomes more significant at later times during the flare decay. We have
also investigated the $f/i$ behavior during the late secondary flare after $0.84$~d, but no
significant density increase as in the earlier episodes could  be observed.

It is interesting to note that the density behavior of this flare finds parallels 
in observations of large solar flares. \citet{mckenzie80} presented O\,{\sc vii} flux
ratios during a large flare with similar time scales as ours and found
$f/i$ ratios around unity close to the flare peak, implying densities
of up to $2\times 10^{11}$~cm$^{-3}$. Much shorter flares were discussed
by \citet{doschek81}; in those cases, the densities reached peaks around $(10-20)\times
10^{11}$~cm$^{-3}$ as measured from O\,{\sc vii}, but this occurred very early in these flares (during
the flare rise), while during flare peak, electron densities of a few times 
$10^{11}$~cm$^{-3}$ were derived. As in our observations,
the densities very rapidly decreased after the peak to levels at or below
$10^{11}$~cm$^{-3}$. The estimated masses and volumes, on the other hand, steadily
increased, a conclusion also drawn in Paper I for Proxima Centauri where we suggested that this 
indicates an increase of the amount of cool plasma that was initially heated  to
higher temperatures but that cools to a few MK. Unfortunately, we have no
reliable density diagnostic for the more relevant flare temperatures. 
\citet{doschek81} and references therein  report solar-flare densities
derived from Fe\,{\sc xxv} ($T > 10$~MK) that are similar to those measured from 
O\,{\sc vii} ($T \approx 2$~MK). \citet{landi03} recently used various density
diagnostics for an M class solar limb flare. From Fe\,{\sc xxi} lines, they derive
densities up to $3\times 10^{12}$~cm$^{-3}$, but there are conflicting
measurements for lower ionization stages that reveal much lower densities, 
comparable with pre-flare densities (see also references in that paper).
If the density values at $\approx 10^7$~K are real, then  pressure equilibrium cannot 
be assumed for flaring loops; a possible explanation involves volumes
for the O\,{\sc vii} and the Fe\,{\sc xxi}/Fe\,{\sc xxv} emitting plasmas that are spatially separate, possibly
contained by different magnetic field lines. Obviously, further solar studies are 
needed.

\subsection{Abundances and emission measure distributions}

A derivation of elemental abundances is inherently related to 
the knowledge of the EM distribution because the
formation efficiency of all observed X-ray emission lines in a collisionally 
ionized plasma is strongly dependent on the temperature structure.
For previous efforts 
using high-resolution X-ray spectroscopy, see, e.g.,
\citet{brinkman01, drake01, guedel01a} and \citet{audard03a}. 
EM  distributions are also discussed in \citet{huenemoerder01}. Despite some
differences between the reported results, there is wide agreement
that magnetically very active stars reveal an {\it inverse First-Ionization
Potential} effect \citep{brinkman01} that expresses itself in low abundances of 
elements with a low first ionization potential (FIP), and enhanced
abundances of elements with a high FIP. This is in contrast to the
Sun \citep{vonsteiger89} and solar-like inactive stars in which a normal FIP effect
has been found in similar X-ray studies (enhanced low-FIP elements; \citealt{guedel02b}).

There is particular interest in studying elemental abundances also
during flares. Flares bring chromospheric and photospheric material
into the corona after heating this cool gas and inducing 
chromospheric evaporation. Differences between  the composition
of flare plasma, chromospheric plasma, and the overall low-level
coronal plasma could have important implications for mass transport
or diffusion processes. While overall abundance changes in 
flares have been reported previously (e.g., \citealt{mewe97}),
a selective FIP-related change of elemental abundances was first
reported for a large flare on an RS CVn-type binary by \citet{guedel99}.
Similar enhancements of low-FIP elements during large
flares on other RS CVn binaries were described by \citet{osten00} and
\citet{audard01}. We will test here 
whether similar enhancements are seen in the Proxima Centauri flare.

Despite the comparatively large count rates in the present flare on 
Proxima Centauri,  the analysis of abundance anomalies and
the differential EM distribution is challenging because
of the short time scales of variability. We again use sections Q and  A--E 
marked in Fig.~\ref{light1} for the analysis.
We  subjected the PN and RGS1+2 data to the same analysis method as
described in \citet{audard03a}. In  short, we used only the strongest
emission lines of the RGS and their immediate surrounding wavelength
ranges, in particular lines of Mg, Fe, Ne, O, N, and C, plus a few
intervals of continuum. We thus preserve a correct
treatment of the overlapping line wings in particular in the Fe L-shell
region while discarding many regions containing weak and ill-defined
emission lines with often poor atomic data. We also flagged wavelength
intervals with poor instrumental calibration, in particular all
data shortward of about 8.3~\AA\ in the RGS. The description of the important hot flare
plasma components requires additional information from the EPIC PN
camera that sensitively constrains the hot end of the EM distribution
by the bremsstrahlung continuum and the Fe~K complex at 6.7~keV.
Since the high-resolution RGS data discriminate much better between
spectral features at lower energies, we used the PN data only above
1.3~keV (shortward of 9.3~\AA). The combined PN + RGS1 + RGS2 
data sets were then subject to
a multi-component analysis in the SPEX software \citep{kaastra96a}
to derive an initial set of temperatures, EMs, and
abundances. This procedure has the advantage of optimally considering
all line blends and wings that may affect the selected line systems,
while at the same time self-consistently calculating the  (abundance and
temperature-dependent) continuum level. The abundances were then used to 
derive a more comprehensive EM distribution. In principle, 
this procedure could be iterated, but the limited data quality at hand
does not warrant a deeper analysis.

\begin{figure} 
\hbox{\includegraphics[width=8.8cm]{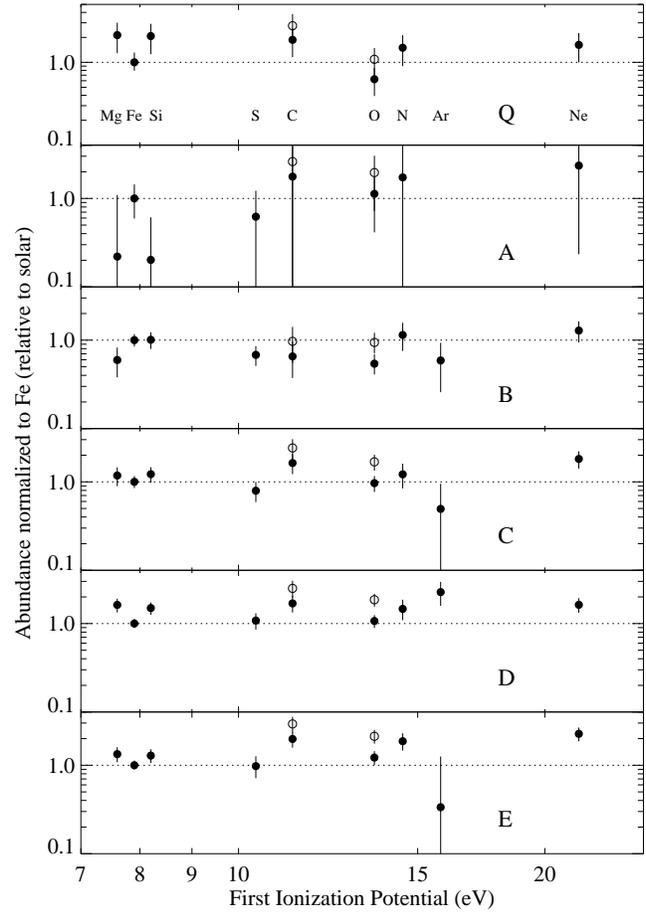}}
\caption{Elemental abundances determined during the low-level episode Q and 
during flare intervals A--E. All abundances are plotted as ratios to the
Fe abundance, and are normalized with the respective solar photospheric abundance
ratios according to \citet{anders89}. For Fe,  an updated value 
given by \citet{grevesse99} has been used. The open circles refer to solar abundances of
O and C reported by \citet{allende01} and \citet{allende02}, respectively.\label{abun}}
\end{figure}

We report here the abundance ratios (with respect to the Fe abundance) and EM distributions
obtained from the intervals Q and A-E. All ratios are normalized to solar
photospheric ratios as given by \citet{anders89}. For Fe alone, we adopted 
an updated solar photospheric abundance reported by \citet{grevesse99}.
We prefer to use these traditional solar values for ease of comparison with previous 
X-ray studies. We do, however, also show the abundance ratios if the
updated solar photospheric values of C and O presented by \citet{allende01, allende02}
are used as a basis. 
The results relative to the former set of solar photospheric abundances
are listed in Table~\ref{abundancetable} together with their 1$\sigma$
rms errors, and illustrated in Fig.~\ref{abun} (filled circles). Since the updated Allende 
Prieto et al. photospheric abundances of C and O are smaller by factors of 1.48 and 1.74, 
respectively, than those given by Anders \& Grevesse, the ratios in Fig.~\ref{abun}
increase by the respective factors. These modified abundance ratios are shown by
open circles.

\begin{figure} 
\hbox{\includegraphics[width=8.8cm]{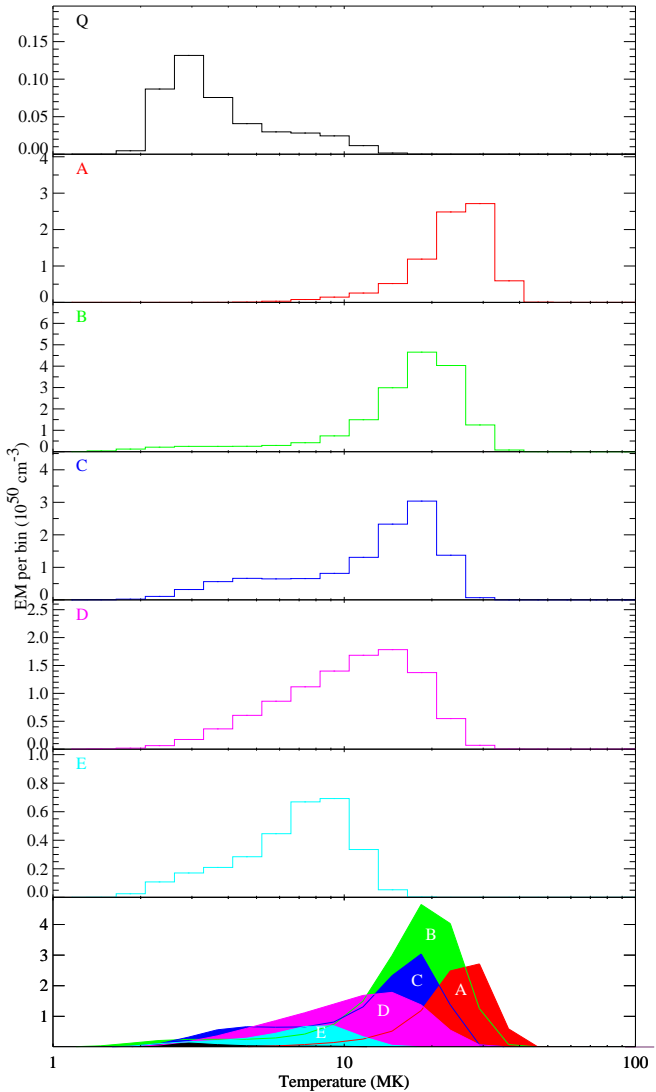}}
\caption{Emission measure distributions for the pre-flare interval Q and for flare intervals A--E,
using a polynomial reconstruction algorithm that fits the DEM with polynomials of order 6.
The lowest panel shows a superposition of all flare DEMs on the same absolute scale.
\label{dempoly}}
\end{figure}

\begin{figure} 
\hbox{\includegraphics[width=8.8cm]{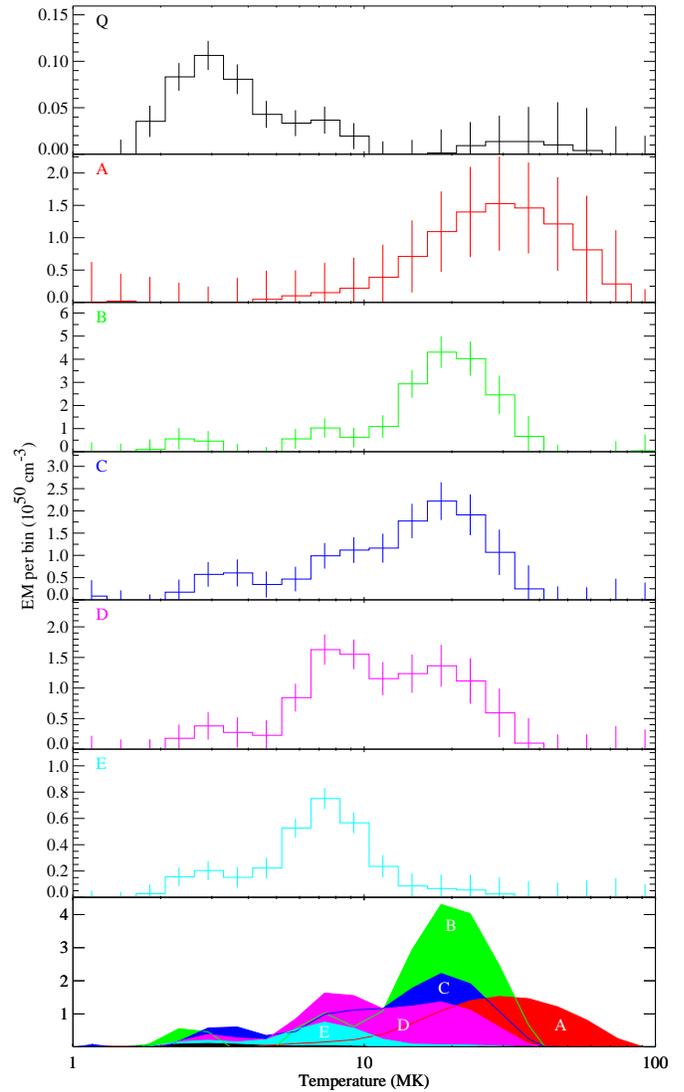}}
\caption{Similar to Fig.~\ref{dempoly}, but using a regularization inversion of 
the spectrum to obtain the EM distribution.\label{demreg}}
\end{figure}

The most surprising aspect here is that the abundance ratios with respect to
Fe are nearly constant across the range of FIP, and most ratios are actually
close to solar photospheric values (ratios of unity). The abundances of Ne, N, and
C tend to systematically exceed unity, although the effect is, given the error bars
and unknown systematic errors in the atomic physics (in particular of several blending
lines), rather small. If the updated solar abundances for C and O given by
\citet{allende01, allende02} are considered, then  there is some trend during 
the late flare phases for high-FIP elements to be higher than low-FIP elements (by factors
of 2--3), while this trend is not visible during flare peak.
We also quote the ``absolute'' abundance (with regard to H) 
of Fe for each interval; such values require good measurements of the continuum which we
feel confident we have well determined using the high-energy tail of the 
EPIC PN spectrum (see Fig.~\ref{epic}), given the high dominant temperatures
in the flare (Figs.~\ref{dempoly}, \ref{demreg}).

Fig.~\ref{dempoly} and \ref{demreg} illustrate two versions of the corresponding
EM distributions. The former set was obtained by fitting 
Chebychev polynomials of order 6, while the latter set uses
an integral inversion technique with a regularization
parameter for the smoothness of the DEM (errors cannot be derived in the context of 
the polynomial fit, but comparing with the alternative regularization method illustrates the 
range of solutions). The differences between the methods are not 
surprising and relate to the ill-posedness of the underlying integral inversion problem.
Details of these methods are described in \citet{kaastra96b}. A comparative
study of several inversion techniques is given in \citet{guedel97a}.
Overall, the polynomial method tends to reveal peaks in the EM
distribution, while the regularization method shows extended
wings up to high temperatures. The two versions illustrate the
principal uncertainty in the reconstruction of an EM
distribution but at the same time clearly illustrate
the main thermal structure of the flare, and its change across
the different flare phases. The lowest panel in each of these figures illustrates
the evolution of the EM distribution by showing all DEMs on the same scale. Although
the regularization method clearly spreads out the EM over a larger range than
the polynomial method, the trends are the same: Rapid heating to high temperatures
is followed by an increase of the EM to its peak at somewhat lower temperatures, 
after which the EM spreads out over a larger range while cooling.


\begin{table*}
\centering
\caption{
Abundance ratios with respect to Fe for various phases (A-E) of the large flare (values rounded to multiples
of 0.1), and
for the low-level pre-flare section (Q). All values are normalized with the 
solar ratios. The basic solar abundances of \citet{anders89} have been used, except 
for Fe for which the value given in \citet{grevesse99} has been adopted. Errors are 
based on 1$\sigma$ errors of the individual abundances (see text for details). ``Fe/Fe''
is unity by definition, but the error bars indicate the fractional uncertainty of 
the Fe abundance.
The last line gives the Fe abundance relative to solar photospheric values.
}
\begin{tabular}{lllllll}
\hline
Element & Q & A & B & C & D & E \\
\hline
C/Fe    & $1.9_{-0.7}^{+0.7}$ & $1.8_{-1.9}^{+3.8}$ & $0.7_{-0.3}^{+0.3}$ & $1.6_{-0.4}^{+0.4}$ & $1.7_{-0.4}^{+0.4}$ & $2.0_{-0.4}^{+0.4}$ \\
N/Fe    & $1.5_{-0.6}^{+0.6}$ & $1.7_{-1.9}^{+5.2}$ & $1.1_{-0.4}^{+0.4}$ & $1.2_{-0.4}^{+0.4}$ & $1.5_{-0.4}^{+0.4}$ & $1.9_{-0.4}^{+0.4}$ \\
O/Fe    & $0.6_{-0.2}^{+0.2}$ & $1.1_{-0.7}^{+0.6}$ & $0.5_{-0.1}^{+0.2}$ & $1.0_{-0.2}^{+0.2}$ & $1.1_{-0.2}^{+0.2}$ & $1.2_{-0.2}^{+0.2}$ \\
Ne/Fe   & $1.6_{-0.6}^{+0.6}$ & $2.4_{-2.1}^{+2.2}$ & $1.3_{-0.3}^{+0.4}$ & $1.8_{-0.4}^{+0.4}$ & $1.6_{-0.3}^{+0.3}$ & $2.3_{-0.4}^{+0.4}$ \\
Mg/Fe   & $2.1_{-0.8}^{+0.9}$ & $0.2_{-0.2}^{+0.9}$ & $0.6_{-0.2}^{+0.2}$ & $1.2_{-0.3}^{+0.3}$ & $1.6_{-0.3}^{+0.3}$ & $1.3_{-0.3}^{+0.3}$ \\
Si/Fe   & $2.1_{-0.8}^{+0.8}$ & $0.2_{-0.2}^{+0.4}$ & $1.0_{-0.2}^{+0.2}$ & $1.2_{-0.3}^{+0.2}$ & $1.5_{-0.3}^{+0.2}$ & $1.3_{-0.2}^{+0.2}$ \\
S/Fe    & $ -               $ & $0.6_{-0.5}^{+0.6}$ & $0.7_{-0.2}^{+0.2}$ & $0.8_{-0.2}^{+0.2}$ & $1.1_{-0.2}^{+0.2}$ & $1.0_{-0.3}^{+0.3}$ \\
Ar/Fe   & $ -               $ & $ -               $ & $0.6_{-0.3}^{+0.3}$ & $0.5_{-0.4}^{+0.5}$ & $2.3_{-0.7}^{+0.7}$ & $0.3_{-0.3}^{+0.9}$ \\
Ca/Fe   & $ -               $ & $ -               $ & $ -                  $ & $ -            $ & $1.5_{-1.0}^{+1.0}$ & $ -	   $ \\
Fe/Fe   & $1.0_{-0.2}^{+0.3}$ & $1.0_{-0.4}^{+0.4}$ & $1.0_{-0.2}^{+0.2}$ & $1.0_{-0.2}^{+0.2}$ & $1.0_{-0.1}^{+0.1}$ & $1.0_{-0.1}^{+0.1}$ \\
Ni/Fe   & $0.8_{-0.8}^{+0.9}$ & $3.1_{-3.1}^{+3.0}$ & $3.1_{-0.9}^{+0.9}$ & $3.6_{-1.0}^{+1.1}$ & $4.1_{-1.0}^{+1.0}$ & $2.7_{-0.8}^{+0.8}$ \\
\hline														     
Fe/H    & $0.51_{-0.11}^{+0.16}$ &$0.58_{-0.24}^{+0.26}$  & $1.21_{-0.19}^{+0.20}$ & $0.67_{-0.10}^{+0.11}$ & $0.54_{-0.06}^{+0.07}$ & $0.47_{-0.05}^{+0.06}$\\
\hline
\end{tabular}
\label{abundancetable}
\normalsize
\end{table*}

\subsection{Optical depth}

Standard interpretation of coronal emission assumes that the optical
depth of the plasma is negligible for both line and continuum emission. Traditional
spectral analysis of coronal EM distributions and elemental 
abundances would be fundamentally flawed if significant optical depths
were present, and radiative transport methods would be in order. \citet{schrijver94}
argued for non-zero optical depth due to resonant scattering in
several lines of the inactive solar analog $\alpha$ Cen  in the extreme ultraviolet range, a view that
was challenged by a follow-up investigation including EM
analysis from X-ray data \citep{schmitt96}. The situation is also
rather unclear in the solar context. While \citet{schmelz97} and \cite{saba99}
find evidence for optical depth effects in the Fe\,{\sc xvii} $\lambda$15.01
line based on flux ratios with respect to the Fe\,{\sc xvii} $\lambda$15.26 and
the Fe\,{\sc xvii} $\lambda$16.78 lines, recent laboratory
measurements  of these ratios significantly differ from previous theoretical 
calculations \citep{brown98, brown01, laming00}.

Stellar coronal optical depths have recently been studied using 
grating observations from {\it XMM-Newton} \citep{audard03a} and {\it Chandra}
\citep{ness01, ness03a}, with results compatible with
negligible optical depths in stars across a wide range of activity levels.
Conditions in large flares may, however, be different. Following
\citet{ness03a}, we measure line ratios between lines with
low oscillator strengths, namely Fe\,{\sc xvii} $\lambda$15.26 and
Fe\,{\sc xvii} $\lambda$16.78, and a line with a high
oscillator strength, namely Fe\,{\sc xvii} $\lambda$15.01. The flux in
the latter is thus more likely to be reduced by resonant scattering. 
Experimentally, for zero optical depth
$f(15.26)/f(15.01) = 0.30-0.36$ (\citealt{ness03a} and references therein). 
For the $f(16.78)/f(15.01)$
ratio, previous theoretical values range between $0.40-0.50$, while
newer calculations by \citet{doron02} give a range between $0.60-0.75$
depending on temperature (see also the detailed description in 
\citealt{ness03a} and references therein).

The S/N ratio of our data was sufficient to derive some rough
values for the above flux ratios. We fitted the lines with delta functions
at their rest wavelengths convolved with the instrument
response function. The continuum level  was described, in the vicinity
of the respective lines, by a power-law function. Since the line wings of
the Fe\,{\sc xvii} $\lambda$15.01 and the Fe\,{\sc xvii} $\lambda$15.26 overlap
and an intervening Fe\,{\sc xix} line is also detected at $\lambda$15.20,
these lines were simultaneously fitted. After a first iteration,
the wavelengths of the brighter two lines (those at 15.01\AA\ and 16.78\AA)
were optimized by using them as fitting parameters as well. The deviations
were minor.

From our data, we find $f(15.26)/f(15.01) = 0.18-0.31$ for the intervals 
C, D, and E where the S/N is sufficient for meaningful ratios, with
an error of about $\pm(0.06-0.09)$ for each ratio. For $f(16.78)/f(15.01)$, we
find, for the same three intervals, values of $0.43-0.62$, with errors
of $\pm(0.06-0.12)$. We thus clearly do not detect 
any opacity effect at any of the flare phases investigated, at least within
the large error ranges allowed by the present data.  This
suggests that even under extreme conditions of  a flare 
plasma, optical depth effects are negligible, at least in the temperature
range investigated with Fe\,{\sc xvii}  lines ($ T \approx 3-8$~MK).

\section{Interpretation and discussion}

\subsection{Small flares and coronal heating}

The first portion of our observation shows continuous variability to
an extent that no part of the observation can be defined as being ``quiescent''
for intervals exceeding $\approx 1$~hr (see Fig.~\ref{light1}), and often variability
occurs on much shorter time scales.
This is a new aspect made possible by {\it XMM-Newton's}  high sensitivity and 
the long uninterrupted monitoring (see also \citealt{audard03b} for a similar
observation of the late-type M dwarf binary UV Cet). We suggest that the numerous small peaks
visible in the X-ray light curve on time scales of 1--10 minutes are the most prominent 
examples of a large sequence of intermediately luminous flares, approximately corresponding
to M class flares in the solar classification. Support for this view is manifold:
i) The light curves of the small enhancements resemble typical flares on active stars
or on the Sun (Paper I for examples). ii) We find many of these enhancements
to be preceded by optical bursts similar to the standard behavior in solar flares, and
also similar to the correlated light curves of the large Proxima Centauri flare.
iii) Hardness increases during periods of enhanced low-level activity.

We suggested, in Paper I, that much of the X-ray emission observed during this observation
may be due to the superposition of a large number of intermediate and weak flares, most
of which cannot be individually resolved, therefore forming a pseudo-continuous 
emission level with sporadic enhancements due to the strongest but least numerous
members of the distribution. The principal model we thus propose is fundamentally 
different from a static corona and is rooted in evidence
that magnetically active stars show mean coronal
properties that are at variance with a corona like the
Sun's. In a comparative study of solar analogs at
different activity levels, \citet{guedel97b} studied stellar
X-ray EM distributions as a function of 
total X-ray luminosity (or age) and compared them with 
time-averaged EM distributions of solar 
flares. They found that active stellar EM distributions resemble the 
superposition of a low-level solar EM distribution and a 
time-averaged EM of strong (solar) flares as interpreted from
GOES satellite data. Toward lower-activity stars, the
hot component becomes weaker, probably due to a smaller
rate of flares heating the corona to high temperatures. 
The complete active-stellar EM distributions may thus be the result of 
stochastic, superimposed flaring \citep{guedel97c} .
In  more active stars, the rate of strong flares is larger, but
since larger flares heat more EM to higher temperatures than
do small flares \citep{feldman95}, the time-averaged EM distribution develops
a prominent high-temperature component additional to the low-T
emission measure. The latter is, in this picture, due to the combined
effect of microflares that heat the plasma to only a
few MK. The physical cause for an increase of the flare rate,
and hence for an increase of the rate of {\it large} flares heating 
a large amount of plasma to high temperatures, was suggested to
be the increasing filling factor of active regions \citep{guedel97b}.
As the concentration of active regions on an active star becomes larger, 
there are more interactions between tangled magnetic fields, leading to 
more numerous flares. The rate of very large flares thus also increases.

This interpretation is also in line with recent studies of longer time series
obtained with the EUVE and BeppoSAX satellites (\citealt{audard00, kashyap02, guedel03a}; see also
\citealt{butler86}) that suggest
that {\it all of} the observed extreme ultraviolet and X-ray emission of active
late-type main-sequence stars could be made up by a population of stochastic flares 
distributed in total energy as a power law, with the smallest flares dominating the energy budget.
All observed X-ray emission characteristics would thus be a consequence of a multitude
of superimposed flare rises and decays. For example, the EM distribution
would be dominated by the statistical properties (temperature, density, volume) 
attained during the long flare decays \citep{guedel97b, guedel03a}.

Proxima Centauri is, during its low-level episodes, a moderately  active star only,
with $L_{\rm X}/L_{\rm bol} \approx (0.6-2.4)\times 10^{-4}$, and 
with an EM distribution dominated by $\approx$3~MK plasma and
a tail reaching to $\approx$10~MK (Figs.~\ref{dempoly}, \ref{demreg}). We
expect large flares to be much hotter. \citet{feldman95} found a correlation
between peak EM and peak temperature of solar and 
stellar flares. The faintest detectable flares in our observations were
estimated to be $(2-5)\times 10^{26}$~erg~s$^{-1}$ (Sect.~\ref{lightc}), which
implies EM $\approx (7-16)\times 10^{48}$~cm$^{-3}$ if a specific plasma emissivity per unit 
EM of  $3\times 10^{-23}$~erg~cm$^{3}$s$^{-1}$ is taken as a basis (from 
the SPEX software for temperatures of a few MK, using solar abundances according to
Table~\ref{abundancetable} for the flare peak; see~\citealt{kaastra96a}).
Fig.~2 in \citet{feldman95} then suggests flare peak temperatures of $15-20$~MK. 
Since these are only the largest of the contributing flares during low-level episodes
and the characteristic temperatures during the long decay are typically
a factor of two lower, our observations are well compatible with the view that 
low-level flares dominate the EM distribution. 
\begin{figure} 
\includegraphics[width=8.8cm]{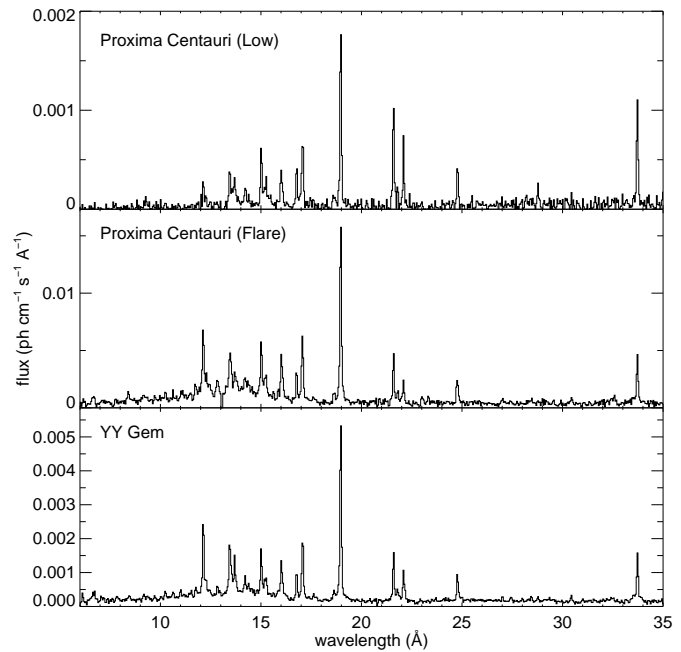}
\caption{Comparison of the fluxed RGS spectrum of YY Gem (bottom panel)
          with the spectrum obtained during the large flare (combined
	  intervals A, B, C, and E; middle panel) 
	  on Proxima Centauri, and the low-level spectrum (Q; top panel). 
	  The spectral bin width is 43.74\AA.\label{yygem}}
\end{figure}

We can illuminate this problem from another angle. Extremely X-ray active M dwarfs
with luminosities much higher than Proxima Centauri show higher characteristic coronal
temperatures during low-level episodes. For example, the DEM of the active dM1e+dM1e binary 
YY Gem is dominated by plasma at $T = 6-12$~MK \citep{guedel01b}. The light curve is,
somewhat similar to Proxima Centauri, interspersed with intermediate and weak flares,
while a quasi-steady emission level dominates the overall spectrum. The  difference is 
that YY Gem's low-level luminosity is two orders of magnitude higher. If the flare heating hypothesis
as described above holds, then the low-level emission of YY Gem is dominated by proportionately
larger flares, namely flares $\approx$100 times stronger than those barely seen during
Proxima Centauri's low-level emission, thus producing higher characteristic temperatures
according to the Feldman et al. relation. 
Incidentally, the large flare on Proxima Centauri described here corresponds to such
an event. We would then expect that the high-luminosity, long decay of the 
Proxima Centauri flare determines a spectrum that resembles the low-level emission of YY Gem. 
This is indeed the case. Fig.~\ref{yygem} compares the fluxed RGS spectra of the combined
flare portions A, B, C, and E (omitting the secondary flare peak)  
with the integrated spectrum of YY Gem collected from a 
$\approx 1$~d observation \citep{guedel01b}. The spectral line ratios and the continuum
level look surprisingly similar but differ significantly from the much cooler spectrum
observed during Proxima Centauri's low level emission. The flare DEM tends to be somewhat hotter 
than the integrated YY Gem DEM \citep{guedel01b}, and the Ne\,{\sc ix} and Ne\,{\sc x} lines at 13.45~\AA\ and
at 12.1~\AA, respectively, tend to be somewhat lower in the Proxima Centauri flare compared to the
Fe\,{\sc xvii} lines, which is probably due to  differences
in the Ne/Fe abundance ratio (Fe\,{\sc xvii} and Ne\,{\sc ix} have very similar line formation temperatures).
Note, however, that the DEM reconstruction presented
by \citet{guedel01b} was based on RGS2 only; this spectrum is rather insensitive to
temperatures significantly above 10~MK. The MOS spectra (Table 2 in \citealt{guedel01b})
indicate an additional large amount of EM around 20~MK (see also
\citealt{stelzer02}).  The presence of weaker flares and later flare decays will also
increase somewhat cooler components in the DEM. This comparison again suggests that more active 
stars are dominated by flare-like plasma corresponding to  intermediate-to-large flares, reaching to high 
temperatures, while lower-activity coronal emission may as well  be composed of flare contributions
which, given the lower flare energies, attain lower average temperatures. This hypothesis 
would explain i) why more active stars show hotter coronae, ii) why more active stars 
show much higher luminosities (due to the increased EM built up during the
flare process). It further supports the model that active stellar coronae are heated by 
stochastic flares. 
And lastly, it may explain why active (main-sequence) stars show
high characteristic densities. If low-level emission is made up of flares as the one discussed
in this paper, then  densities derived spectroscopically from the time-integrated spectrum across
the entire flare should be similar to densities measured in active stars during quiescence.
For the Proxima Centauri  flare, we obtained an average characteristic density from the
O\,{\sc vii} triplet of log$n_e = 10.50^{+0.32}_{-0.25}$, while, for example, the low-level YY Gem spectrum 
reveals a similar value, namely log$n_e = 10.35^{+0.13}_{-0.45}$, as
do many other active main-sequence stars \citep{guedel01a, ness02}.

\subsection{Flare models} 

We will now discuss the large flare in the context of standard flare
models.  Various models have been discussed in the literature,
although all of them  represent simplified paradigms that may not
be realized in the Sun or stars in their pure forms. Rather, we expect
that some gross characteristics of the models may approximately
describe the observations, thus providing some insight into 
the overall development of magnetic energy release. 
With this in mind, we will, in the following, describe
two complementary models and investigate their applicability
to large flares as the one presented here. The first approach, a
2-ribbon flare model based on work by \citet{kopp84} but extended
to include approximate radiative and conductive cooling losses, was 
presented in this form by \citet{guedel99}. The second approach, 
using full hydrodynamic simulations of a closed loop or a pair of
loops, has been previously described by, e.g., \citet{peres82}.
We will discuss one typical numerical solution, but will
defer an in-depth presentation of the hydrodynamic simulations to a companion paper 
\citep{reale04}.  We begin this discussion by briefly reviewing some 
information on the initial energy release phase that can be extracted 
from the optical burst. 
 
\subsubsection{The impulsive phase of the flares} 

The initial energy release of a solar flare is in general most obviously seen
at radio wavelengths, in the optical/UV regime, and in non-thermal
hard X-rays. The relevant signals in these three wavelength regimes are
thought to be related to high-energy electrons accelerated in the initial
phase of a flare in the coronal reconnection region. While radio signals
are due to gyrosynchrotron emission of electrons spiralling in
the magnetic fields, the hard X-ray signal is due to bremsstrahlung when a
non-thermal population of electrons impacts in the chromospheric/transition region 
layers at the magnetic field footpoints. At the same location, optical emission
is thought to arise promptly due to heating and ionization of the chromospheric
plasma. As this outline suggests, the three signatures of the impulsive flare
phase should occur early in the flare, as the electrons are thought to
be the cause for the subsequent chromospheric evaporation that then dominates
the soft X-ray emission \citep{antonucci84, dennis88}. For the optical emission
observed in Proxima Centauri, this is clearly the case as was shown in Paper I
(see also Fig.~\ref{light2}). Furthermore, solar observations clearly show - as
expected - a very close time correlation between the hard X-rays and
the microwave emission \citep{kosugi88}, and between optical ``white light'' 
and hard X-ray emission (\citealt{hudson92}; \citealt{dennis88} and 
references therein), often within seconds.

While solar research has made extensive use of non-thermal hard X-rays for
the study of the energetics of the impulsive flare phase \citep{dennis88},
the respective signals are presently out of reach for stellar astronomy
since required detector sensitivities are not available at the energies
of interest ($>$20~keV). However, observation of ``white light'' flares (or 
equivalently, as in this work) U band flares is relatively easy for M dwarfs 
given the large contrast of flares against photospheric background emission.
The U band has traditionally played a major role since the flare enhancement
is particularly large there. The physics of the continuum enhancement
is not very well understood. \citet{hudson92} (and references therein) 
argue for excitation by energy transfer from electron beams, either by  
direct collisions, or by UV irradiation excited by the electron impact. For a further,
short discussion on the issue, see \citet{hawley95}.

Independent of the poorly understood physical cause of the U band emission,
our observation, together with previous solar knowledge puts us into a position
to at least grossly assess the time evolution of the primary energy release
in electrons if the standard model outlined above holds. The U-band light 
curve (above its photospheric level) would thus be a sensitive tracer of the 
production of non-thermal particles in the corona, and therefore {\it an
efficient proxy of the inaccessible non-thermal hard X-ray emission} so often
seen in solar flares.

A second application is possible, namely to roughly determine  the chromospheric
footpoint area of the flaring magnetic fields, as outlined in \citet{hawley95}.
To do this, one would have to know the temperature of the U-band emitting
source,  requiring at least one more observation in another optical band, which is not 
available. The analysis by \citet{hawley95} for flares on the M dwarf AD Leo
suggests temperatures of $\approx 9000-9500$~K under the assumption  of blackbody 
emission. Assume that the filling factor of the flare footpoints in the projection onto
the sky is $X$. Since the area emitting at the non-flaring photospheric
temperature is reduced by the flaring area, the  U-band luminosity ratio between 
measurements taken during the flare  and outside the flare is   
\begin{equation}
R = {L_f + (1-X)L_q\over L_q} = {XB_{\rm U}(T_f) + (1-X)B_{\rm U}(T_q)\over  B_{\rm U}(T_q)}.
\end{equation}
Here, $L_f$ is the U-band flare luminosity, and $L_q$ is the U band quiescent  luminosity
of the whole star; $B_{\rm U}$ is the Planck function evaluated in the U band ($\approx 3500$\AA)
for the flare temperature ($T_f = 9000$~K, see above) and for the photospheric effective 
temperature outside flares ($T_q$ = 2700~K, \citealt{frogel72}), respectively. Therefore, recalling
that $B_{\rm U}(T) \propto (\mathrm{exp}(hc/k\lambda T) -1)^{-1} \approx \mathrm{exp}(-41156/T)$, we find 
\begin{equation}
R = 1 + X\left( e^{41156(1/T_q - 1/T_f)} - 1\right) \approx 1+4.3\times 10^4X.
\end{equation}
Our flare increased the U band flux by a factor of $R\approx 20$, so that the fractional
projected area of the footpoints is $\approx 4\times 10^{-4}$ of the cross section of the star.
If this described one circular footpoint, then its diameter would be 2\% of the
stellar diameter. Comparing with the results we will derive from 
X-rays below, the area covered by U-band emitting footpoints must be considerably
smaller than the X-ray emitting area, which is compatible with magnetic loops that 
span  over a wide range across the stellar surface.

\subsubsection{A two-ribbon flare model}\label{tworibbon}
 
The two-ribbon (2-R) flare model devised initially by \citet{kopp84}
is intrinsically a parameterized magnetic-energy release model
in which the time development of the flare light curve is fully determined
by the amount of energy available in non-potential magnetic fields, and
the rate of energy release as a function of time and geometry.
The complete time evolution of the flare is based on
continuous heating that feeds an evolving reservoir of 
hot plasma. Plasma cooling is not included in the original model; it is assumed
that a portion of the total energy is radiated into the observed X-ray band,
while the remaining energy will be lost by other mechanisms.  
2-R flares are well established for the Sun; they often lead to large, long 
duration flares that may be accompanied by mass ejections. This configuration 
is clearly suggestive for our observation, paralleling conclusions 
for similar flares observed previously on Proxima Centauri \citep{haisch83}.
We will describe here an extension of the model that includes
approximations of radiative and conductive losses, and that will also
take into account some geometric development of the radiating magnetic structures.
This variant of the 2-R model was first applied to a large, long-duration
flare observed on the RS CVn-type binary UX Ari \citep{guedel99}. A detailed description
is given in the latter reference; we give a brief summary below.
 
A 2-R flare conceptually starts when a  disruptive event,
e.g., a filament eruption, opens a loop arcade. The open field lines are then
driven toward a radial neutral sheet (above the magnetic neutral line) where 
they reconnect at progressively higher altitudes, forming a system of
overlying loops below them. The excess energy  provided by annihilation
of the magnetic fields is released in an unspecified form. The amount of
continuous heating  provided throughout the reconnection episode principally
describes both the flare increase and its decay. 

The magnetic fields are, for convenience, described along meridional
planes on the star by Legendre polynomials $P_n$ of order $n$, up to the height of the
neutral point; above this level, the $B$ field is directed radially. The 
magnetic fields are thus assumed to be poloidal, initially directed radially outward
and then reconnecting, starting with the antiparallel field lines closest to
the neutral line. As time proceeds, field lines further away from the 
neutral line move inward at coronal levels and reconnect at progressively
larger heights above the neutral line. The reconnection point thus moves upward
as the flare proceeds, leaving closed magnetic loop systems underneath. 
One loop arcade thus corresponds to one (N-S aligned) lobe between two zeros of $P_n$ in 
latitude, axisymmetrically continued over some longitude in E--W direction.
The propagation of the neutral point in height, $y(t)$, with a 
time constant $t_0$ (in units of $R_*$, measured from the star's center) 
is prescribed by
\begin{equation}\label{yt}
y(t) = 1 + {H_{\rm m}\over R_*}\left(1- e^{-t/t_0}\right)
\end{equation}
\begin{equation}\label{ht}
H(t) \equiv [y(t) - 1]R_*
\end{equation}
and the total energy release of the reconnecting arcade per radian in longitude is
equal to the magnetic energy lost by reconnection,
\begin{equation}
{{\rm d}E\over {\rm d}y} = {1\over 8\pi}2n(n+1)(2n+1)^2R_*^3B^2I_{12}(n)
        { y^{2n}(y^{2n+1} - 1)\over [n+(n+1)y^{2n+1}]^3} 
\end{equation}
\begin{equation}\label{dedt}
{{\rm d}E\over {\rm d}t} ={{\rm d}E\over {\rm d}y}{{\rm d}y\over {\rm d}t} 
\end{equation}
(Poletto et al. 1988). In Eq.~(\ref{yt}), $H_{\rm m}$ is the maximum height of 
the neutral point for $t\rightarrow \infty$; typically, $H_{\rm m}$  is assumed 
to be equal to the latitudinal extent of the loops, i.e.,
\begin{equation}\label{height}
H_{\rm m} \approx {\pi\over n+1/2}R_*
\end{equation}
for $n > 2$ and $H_{\rm m} = (\pi/2)R_*$ for $n = 2$.
Here, $B$ is the  surface magnetic field strength at the axis of symmetry,
and $R_*$ is the stellar radius. Finally, $I_{12}(n)$ corresponds to 
$\int [P_n({\rm cos}\theta)]^2d({\rm cos}\theta)$ evaluated between the
latitudinal borders  of the lobe (zeros of ${\rm d}P_n/{\rm d}\theta$), 
and $\theta$ is the  co-latitude. We have evaluated
this expression  for lobes that are centered 
at the star's equator for odd $n$, and lobes that start at the equator
for even $n$. Note that $R_*$, $B^2$, and $I_{12}(n)$  
merely define the normalization of the energy release light curve, not its
form. The latter is determined by the time constant $t_0$ and the polynomial
order $n$. 

The largest realistic two-ribbon flare model is based on the 
Legendre polynomial of degree $n = 2$; the loop arcade then stretches out 
between the equator and the stellar poles. For simplicity,
we assume that the arcade extends along a great circle on the stellar surface,
e.g., the equator (instead of adopting axisymmetric arcades at any other
stellar latitude), although the computations of the latitudinal
and radial geometry use the exact equations as given by \citet{kopp84}.
Since  solar two-ribbon flares occur in  loop
arcades whose length  is typically  about 1.5 times their width \citep{poletto88},
we adopt this ratio for any given $n$:
$L = 1.5H_{\rm m} = 3R_*\pi/(2n+1)$.

The Kopp \& Poletto model is applicable 
after the initial flare trigger mechanism has terminated, although 
\citet{pneuman82} suggested   that
 reconnection may start in the earliest phase of loop
structure development. The time of strong production of non-thermal 
electrons, signified by the optical burst during the X-ray increase,
gives rise to explosive chromospheric evaporation and steeply increasing 
temperatures. This phase of the flare may be less well described
by the model. 

\citet{poletto88} introduce an ad hoc factor of 
10\% of the magnetically released radiative energy that  escapes into 
the X-ray regime.  Here, we   
include time-dependent conductive and radiative losses in the X-ray corona 
self-consistently  and hence account for the emitting volume. 
The volume and  the observed EM
provide an overall electron density for which we obtained
estimates from the He-like line triplets in the RGS spectra. Requiring
that the densities agree with the measured densities will thus constrain
the model parameters, in  particular $n$. The radiative loss time is
\begin{equation}
\tau_{\rm r} = {3kT\over n_{\rm e}\psi}
\end{equation}
where $\psi$ is the emissivity per unit EM of the hot, optically thin plasma, which we take
to be $\psi = 2.5\times 10^{-23}$~ergs~s$^{-1}$cm${^3}$ for $T = 15-20$~MK
(from the  collisional ionization equilibrium model in the SPEX software, see 
\citealt{kaastra96a}). The downward conductive flux in a constant
cross section loop is
\begin{equation}
F_{\rm c} = -\kappa T^{5/2}{\partial T\over \partial s}
\end{equation}
with $s$   being the length parameter along the loop, and $\kappa\approx
10^{-6}$~ergs~cm$^{-1}$s$^{-1}$K$^{-7/2}$ \citep{spitzer62}.  $F_{\rm c}$ can be 
approximated as
\begin{equation}
F_{\rm c} \approx -\eta\kappa{T_{\rm m}^{7/2}\over l},
\end{equation}
where the peak temperature $T_{\rm m} \approx T$,  $\eta = 4/7$ has been introduced 
by \citet{kopp93} to account for the geometry in this approximation, 
and $l$ is the time-dependent semi-length of the loop,  
\begin{equation}
l(t) \approx {\pi\over 2}H(t).
\end{equation}
Cooling is principally  controlled by radiation and conduction
with time constants of $\tau_{\rm r}$ and $\tau_c$, respectively, leading to an
effective cooling time  $\tau_{\rm e}$ given by
\begin{equation}\label{tau}
{1\over \tau_{\rm e}}  = {1\over \tau_{\rm c}} + {1\over \tau_{\rm r}}.
\end{equation}
Combining 
the relevant equations (see \citealt{guedel99} for details) one obtains 
for the  model X-ray luminosity
\begin{equation}\label{modellum}
L_{\rm X}(t) = f(t){{\rm d}E\over {\rm d}t}{L\over R_*}
\end{equation}
and the underlying EM must be
\begin{equation}
{\rm EM}(t) = f(t){{\rm d}E/{\rm d}t\over \psi(T)}{L\over R_*}.
\end{equation}
Here, $f(t) = \tau_{\rm e}(t)/\tau_{\rm r}(t)$ gives  the fraction of 
the total energy  being emitted as X-rays, and
$L/R_*$ denotes the length of the arcade in radians, here taken to 
be $\leq 2$ ($L = 1.5H_{\rm m}$).
We  adopt $T$ from the observations, $T\approx 20$~MK during the 
period of consideration. Note
that the {\it observable} X-ray emission is not proportional 
to ${\rm d}E/{\rm d}t$ in Eq.~(\ref{modellum}) as the fraction $f$ 
is time dependent. 
The run of the electron density $n_{\rm e}$ then is
\begin{equation}\label{modeldens}
n_{\rm e}(t) =  {{\rm d}E/{\rm d}y\over 3kTR_*\pi H(t)}.
\end{equation}
The electron density is solely determined by the geometry of the arcade,
as prescribed by Eq.~(\ref{yt})--(\ref{height}).

We integrated all relevant equations in short time steps for a given $n$, adjusting 
$t_0$ and $B$ until a good fit to the observations was obtained. 
Following \citet{guedel99}, we assume that only a fraction  $q < 1$ 
of the liberated magnetic energy is used to heat the plasma (energy that is
lost via radiation and conduction, while the
remaining fraction is transformed into mechanical energy, into fast particles ejected 
from  the corona, etc). \citet{kopp84} use $q = 0.003$ in their
application to a solar flare. 

As the overlying magnetic field lines reconnect, the  loops already reconnected
are pushed downward until they asymptotically reach a final position, corresponding to
a potential field configuration. We have assumed that the 
X-ray loops have relaxed to this final geometry while radiating most
of their X-ray energy (corresponding to  $t\rightarrow
\infty$). The derivation of the asymptotic loop configuration is
performed by piecewise integration as described in \citet{guedel99}.

\begin{figure} 
\hbox{\hskip 0.4truecm\includegraphics[width=8.5cm]{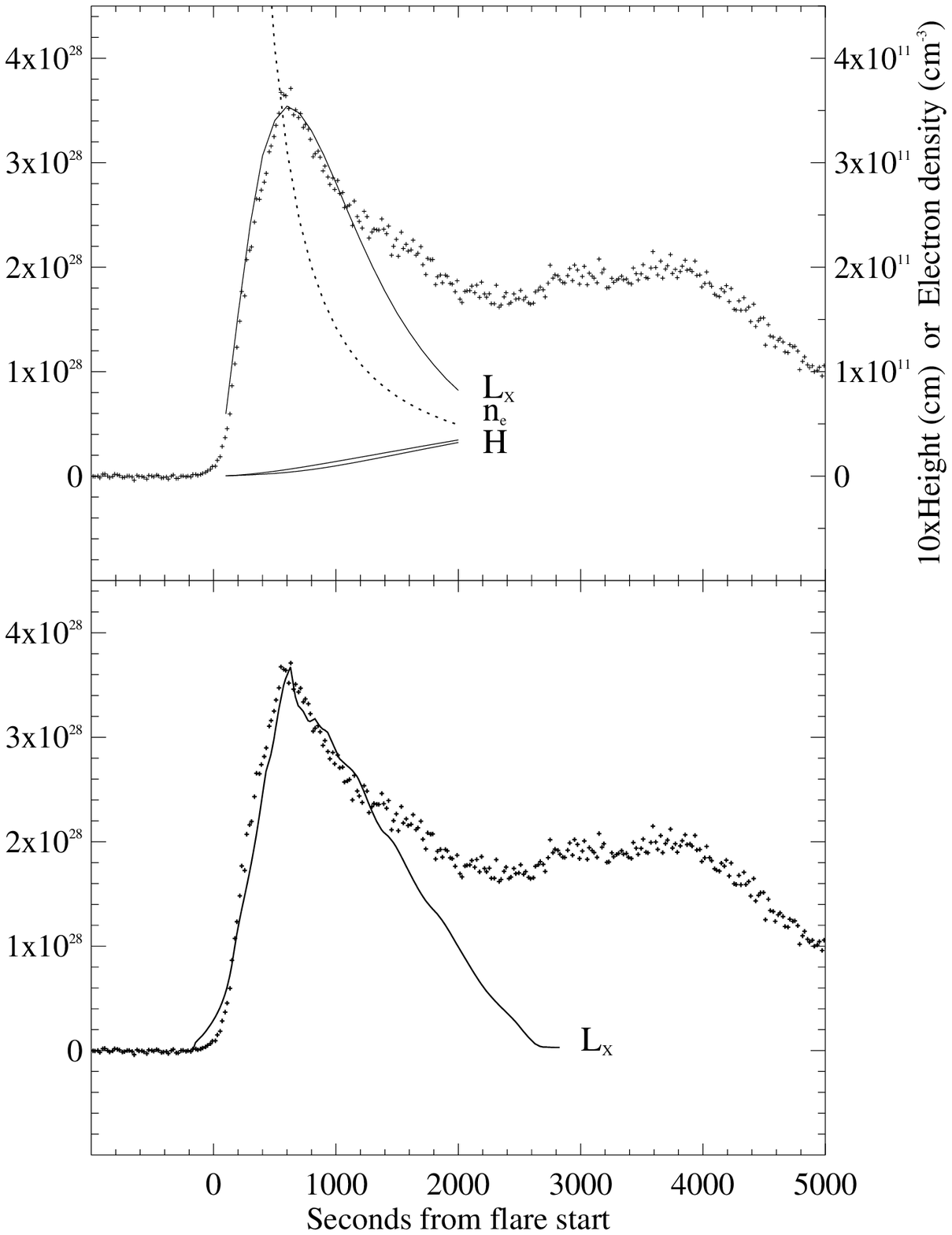}}
\caption{{\bf Top:} Simulation of the large Proxima Centauri flare using a simple 2-R flare
model described in the text. The best fit shown here is for  $n=2$, $B=2050$, $t_0=4000$, $q=0.01$.
The solid line shows the model X-ray luminosity, to be compared with the observations that
are plotted by small crosses. The dotted curve indicates the electron density (see scale on the right),
and the two solid lines at the bottom of the figure mark the height of the emitting loop (see scale
on the right).
The model clearly decays too fast beyond $t = 1000$~s after flare start, which is due to the emergence
of a secondary flare. -
{\bf Bottom:} Similar to upper figure, but here the light curve synthesized from a time-dependent 
hydrodynamic simulation of plasma confined in a single footpoint-heated loop is overplotted
(solid line). The heating lasts 600~s. The flare start is defined slightly differently to
optimize the fit to the rise phase (the precise start of the flare cannot be defined from
the soft X-rays unambiguously). Further details are described in \citet{reale04}.
\label{simul}}
\end{figure}

There are only few independent parameters for this model.
The total energy released is controlled by $B^2$, and by the efficiency $q$.
To obtain the same thermal energy, $q$ obviously scales as
$q \propto B^{-2}$.
The fraction  of radiated X-ray energy out of the total thermal energy is 
given by $f(t)$. These factors together thus determine the X-ray flare 
amplitude. The time scale is given by the parameter $t_0$ in Eq.~(\ref{yt}).
Additionally, Eq.~(\ref{modeldens}) provides an approximation of the loop density
that can be  compared with the observations. 
There is thus a family of solutions $(B, q)$ for each $n$.
The best-fit solution for $n=2$ is shown in Fig.~\ref{simul} (top) and 
required  a maximum surface  magnetic field strength of $B= 2050$~G
for $q = 0.01$. A larger efficiency of $q = 0.05$ thus corresponds to $B = 920$~G.
These two examples characterize the probable range for $B$ although
stronger fields have also been measured in late-type dwarfs \citep{krull00}. 
The time constant $t_0 \approx 
4000$~s, and the final height of the neutral point (for $t \rightarrow\infty$) is 
$H_{\rm m} \approx 1.73\times 10^{10}$~cm or 1.7$R_*$ for $n=2$.

For $n = 3$, we find  $B = 2720$~G for $q = 0.01$, and $t_0 = 3300$~s.
Similarly, for $n = 4$, the best-fit parameters are, for $q = 0.02$: $B = 2200$~G
and $t_0$ = 3100~s. Since we can also estimate the model densities from
Eq.~(\ref{modeldens}), our observations of density-sensitive spectral lines,
in particular O\,{\sc vii}, provide a further constraint. For $n = 2$, Fig.~\ref{simul} shows 
an electron density at flare peak of $n_{\rm e} \approx 3\times 10^{11}$~cm$^{-3}$,
quite compatible with the O\,{\sc vii} measurements. The latter refer to much cooler plasma than
the bulk plasma in a flaring loop at its peak luminosity. However, as discussed
in Sect.~\ref{densovii}, densities inferred for hot solar flare plasma seem to be
of similar magnitude as densities derived from the O\,{\sc vii} lines. 
For $n = 3$, densities at flare peak are $\approx 5\times 10^{11}$~cm$^{-3}$,
while for $n = 4$, $n_e \approx 7\times 10^{11}$~cm$^{-3}$. In the latter case, the
magnetic field estimated from the model at the loop top and at the time of flare peak 
is below the pressure equilibrium value if the electron temperature is $\approx 20$~MK.
We therefore tend to exclude solutions for $n = 4$ and higher within the context of
this simple model.

The model fits well until about $t = 1000$~s after flare start;
thereafter, the model light curve clearly decays too fast, which is due to the emergence
of the secondary flare. Indeed, at $t\approx 1000$~s, the hardness ceases to decay and
develops a plateau (see Fig.~\ref{light2}), presumably due to resumed  heating
\citep{reale04}. We 
emphasize that in the framework of this model, the flare density is not a consequence of 
cooling material condensing out of the corona and
falling back to the surface. Rather, the loops radiating at later times 
are less dense than those radiating at earlier times, according to Eq.~(\ref{modeldens}). 
The 2-R model does not describe the evolution of plasma within a loop, but
rather the plasma and geometry parameters of a series of loops 
in time.

We have to check the value of the decay time $\tau_{\rm e}$. The  
model presented here assumes
that the decay time $\tau_{\rm e}$ is short compared with the evolution of
the flare. If this were not the case, then a significant portion of the
energy released at earlier times would contribute to the emission at
time $t$, i.e., the light curve would be a convolution with radiation
from previous times. From Fig.~\ref{light2}, we estimate that changes in the light
curve occur on a time scale of about 1000~s near flare peak.  The fit
is satisfactory if $\tau_{\rm e, r} \la 1000$~s up to the flare peak. 
Our evaluation of Eqn.~(\ref{tau}) indicates  that $\tau_{\rm e} = 80-100$~s  
around the flare peak for the acceptable solution presented in Fig.~\ref{simul},
and the radiative loss time $\tau_{\rm r}$ is around 800--1000~s,   
which is compatible with our requirements.

We caution that the assumptions for this flare model are crude. 
The purpose of our estimates, however, is to indicate whether
the observed flare belongs to a class of compact flares or 
may be a global phenomenon, compatible with the standard two-ribbon
flare model.  Our results  thus
suggest that the  flare is likely to be geometrically
large, with size scales (height, footpoint separation) on the order of 
$10^{10}$~cm or one stellar radius. 
The presence of such extended 
coronal structures is  compatible with VLBI observations that show flaring 
cores that rapidly evolve into extended structures \citep{mutel85}.
Observed radio sources reveal size   
scales on the order of the stellar diameters \citep{benz98}. Also, the absence of 
radio eclipses or strong signatures of rotational modulation in long-duration flares
on similar M dwarfs agrees well with the assumption of extended
flare geometries on such stars \citep{alef97}.

\subsubsection{Hydrodynamic flaring loop model}

X-ray flares can alternatively be  modeled using full numerical
hydrodynamic simulations. We outline the method here,
but for a technical description, we refer to \citet{peres82} 
and \citet{reale04}. The latter reference discusses applications to
the present observation in deeper detail. The numerical approach stresses 
on plasma confinement. The basic assumption becomes that the flare is mostly
governed by the evolution of the plasma confined in a magnetic loop and
that changes of the loop system geometry and size are small, slow and
play a minor role.

One can then focus on modeling the flaring plasma confined in a single
unchanging loop, where it moves and and transports energy along the
magnetic field lines.  The magnetic field does not play any role
directly, except for plasma confinement.

The confined plasma behaves as a compressible fluid and its evolution
can be described with a one-dimensional time-dependent hydrodynamic
model. When solving the equations of conservation of mass, momentum and
energy, in coronal plasma conditions, several physical terms are
important, and, in particular, the pressure gradients, the stellar
gravity, the radiative losses and the thermal conduction. Describing
all of them requires the numerical solution of the hydrodynamic equations. A
parameterized and time-dependent heating function is included to
trigger the flare. A sufficiently intense heat pulse is able to make
the plasma temperature increase to flare values and to drive a strong
plasma evaporation from the chromosphere: the combination of these two
effects switches the flare on. The flare decay is instead mostly driven
by the plasma cooling by radiation and thermal conduction, after the
heating pulse has finished, unless a residual heating is able to
sustain the decay. This makes an important conceptual difference from
standard 2-R models, in which the decay is assumed to be totally driven
by the release of magnetic energy during the reconnection of
progressively larger arches.

Several numerical time-dependent loop models have been applied to study
solar coronal flares. One of them has been applied to study part of a
flare observed on Proxima Centauri with the Einstein satellite 
\citep{reale88}.

In an accompanying paper we model the flare for most of its duration,
including the second peak, by describing the evolution of flaring
plasma confined in single loop structures. The detailed numerical
approach allows us to synthesize the plasma X-ray emission at the focal
plane of the EPIC PN and to make detailed comparisons with the real
data. We present the overall results of a simulation here for comparison
with our analytic approach using the 2-R formulation.

In the simulations, the flaring plasma confined in a single loop is able to
explain well the rise phase, main peak and initial decay of the flare
(Fig.~\ref{simul}, bottom), and the addition of an arcade of similar flaring loops
allows to explain the second peak and the late decay. In spite of the
different approach, the results are not in contrast to those obtained
with the 2-R modeling described in Sect.~\ref{tworibbon}. Hydrodynamic modeling
and detailed comparison of results to data will however allow us to
obtain specific constraints on the loop geometry and on the heating
function throughout the flare, and insight in the physical flaring
plasma conditions. In particular, the loop height appropriate for
the simulations is very similar to the typical height found in the
2-R model, namely $\approx 0.6\times 10^{10}$~cm. This result is thus in good agreement
with estimates based on the 2-R model. The principal finding from the
modeling is thus that the flaring source is intrinsically large, of 
order of one stellar radius.

\section{Conclusions}

Proxima Centauri is the closest star to the Sun, and it belongs
to the important class of magnetically active stars as well.
The latter are generally characterized by enhanced spot
coverage indicating a large surface magnetic filling factor,
chromospheric emission lines such as H$\alpha$, enhanced flaring 
activity, or strong coronal emissions. Proxima Centauri does not
show signatures of extreme activity but given its surface
area of about 2\% the Sun's, its X-ray luminosity equaling
an average solar level is notable. A more important aspect of
the present and all previous X-ray investigations of this star is 
in fact simply its very late spectral type, which has two
important implications : First, since optical flare emission is
predominantly ``blue'', a large contrast against the weak, red
photospheric emission makes optical flare detection easy. Our
OM data have been an important guide to recognize low-level
flaring in the X-ray light curve. And second, there is quite limited 
surface area to host active regions of a size that would be typical
for the Sun. In fact, Proxima Centauri is no larger than a large
solar coronal active region \citep{reale04}. This circumstance
makes X-ray observations very sensitive to reconfigurations in
the corona that likely affect a considerable portion of the surface.
This obviously helped set a good contrast for the large number of
weak flares we have detected during the low-level X-ray emission.

Our observation has captured a continuously variable X-ray corona,
with flares spanning two orders of magnitude in peak flux and
at least three orders of magnitude in total X-ray energy. The weakest
flares seen during the low-level episode are principally the weakest
of their kind observed on any active star, while the large flare
is the largest yet seen on Proxima Centauri, exceeding even 
some of the strongest solar X-ray flares. Despite this large range of 
flare parameters, we find similar timing characteristics among them:
Many X-ray flares are preceded by optical bursts, suggesting the
process of chromospheric evaporation by electron bombardment as the
main driver of heating and mass transport into the corona. The large
flare is only set apart by its very slow decay and secondary peaks
that indicate continued heating. 

We interpret the low-level emission as being predominantly due to
a superposition of stochastic flaring. While a truly quiescent
emission level due to steadily heated coronal loops cannot be excluded,
the continuous decay of emission by a factor of ten during the first 
12 hours of our observation requires relatively rapid reconfiguration
of the corona unless the complete flux decay is due to the late decay
of a large flare that reached its peak before our observation started.
(Rotational modulation of active regions is not likely to be important
on the observed time scales; the rotation period of Proxima Centauri
is suspected to be as long as 83.5~d, see \citealt{benedict98}.)
The additional modulation on time scales of hours
and associated hardness changes argue for physical changes in the
corona on such time scales. If most of the low-level emission originates 
from flaring plasma, then the flare peaks individually detected
constitute the strongest examples  of a population of low-level 
flares. As such, they determine the high-temperature cutoff of 
the emission measure distribution since flare peak temperatures
are dependent on the peak emission measure \citep{feldman95}. We have
argued that the expected temperatures for these flares are roughly 
commensurate with the observed EM distribution: the DEM shows a high-T 
extension to $\approx 15$~MK, which is the expected peak temperature
for the largest flare during the low-level emission.

If this picture has some merit, then it should apply to other stars at
higher activity levels. For a characteristic distribution of
flare energies as seen on the Sun, higher flare rates also induce
larger rates of larger flares. Therefore, a larger flare rate
(above a certain threshold energy) 
brings more material into the corona and at the same time heats some 
of it to much higher temperatures according to the Feldman et al. 
relation. We thus expect characteristic temperatures in more active stars 
to be higher, which is generally the case. We have scaled this 
model to the extremely active dMe binary YY Gem, which shows an
X-ray luminosity $\approx$100-200 times higher than Proxima Centauri.
The largest flares frequently dominating the low-level emission 
of YY Gem are consequently similar in energy to the large flare
in Proxima Centauri that we captured during the second half of our
observation. The combined spectrum of the latter flare indeed 
resembles the average spectrum of YY Gem, and so do the EM
distributions although a full modeling should take into account 
the additional lower-energy flares occurring on YY Gem as well.

{\it We therefore suggest that the low-level emission of magnetically active
stars is made up, to a large fraction, by a superposition of 
flares over a large range of total energies.} The flare rate (above a
given low-energy  threshold) determines the high-temperature
end of the EM distribution of the intervals
perceived as ``low-level''. A natural consequence of this model is
that, as the flare rate increases toward more active stars, the
emission measure increases because more chromospheric evaporation
is taking place, while the large amount of EM produced by
larger flares is also heated to higher temperatures, thus 
inducing a hotter EM distribution in these stars. {\it We thus suggest
that the \citet{feldman95} relation naturally explains why stars
at a higher activity level show hotter coronae.}

The large flare observed during the second part of our observations
has offered a number of new views on extremely large stellar flares.
Like most of the other, low-level flares, the earliest phase of the rapidly
heating X-ray episode is accompanied by a very large optical burst. 
From solar analogy, we interpret this ``white light'' flare as being 
the  signature of accelerated electrons bombarding the chromosphere
and evaporating the material that is subsequently becoming visible in X-rays.
In this sense, we find that {\it U band flares are ideal and sensitive 
proxies for the non-thermal hard X-rays ($>20$~keV) often investigated 
in solar flare research}. 

While the emission mechanism of the optical flare is somewhat uncertain,
we infer relatively small areas for the chromospheric footpoints.
This is in contrast to the X-ray geometry we find from two different
approaches (an analytic 2-R model and a fully hydrodynamic approach).
Both X-ray flare models indicate relatively large flares, with typical
size scales of $10^{10}$~cm or one stellar radius. While Proxima
Centauri is much smaller than the Sun, the size of such active regions is 
comparable to those that produce very large flares on the Sun. 

For the first time, X-ray spectroscopic measurements have fully resolved the
electron density development during a stellar  flare, indicating  peak
densities in the $1-5$~MK plasma of approximately $4 \times 10^{11}$~cm$^{-3}$.
Such values are similar to densities measured during solar flares.
We have also measured the abundance development during several stages of
the flare. Apart from a trend toward higher abundances of high-FIP elements,
the abundance ratios are relatively close to solar photospheric ratios.

\begin{acknowledgements}
The authors thank Louise Harra for information on solar flare densities.
MG and MA acknowledge support from the Swiss National Science Foundation
(grant  2000-058827, and fellowship 81EZ-67388 to MA), from the Swiss Academy of 
Natural Sciences, and from the Swiss Commission for Space Research. FR acknowledges 
support from Agenzia Spaziale Italiana and Ministero dell'Universit\`a e della Ricerca 
Scientifica e Tecnologica. This research made use of the SIMBAD database,
operated by CDS, Strasbourg. 
\end{acknowledgements}

\end{document}